\documentclass[acmsmall,screen]{acmart}
\usepackage{graphicx} %

\newcommand{\toolname}{\textit{AC4A}}

\usepackage{xcolor}
\usepackage{soul}

\usepackage[frozencache, cachedir=minted]{minted}
\usepackage{tcolorbox}
\usepackage{caption}
\usepackage{hyperref} %
\usepackage{float}    %
\usepackage{multirow}
\usepackage{multicol}
\usepackage{makecell}
\usepackage{booktabs}
\usepackage{colortbl}
\usepackage{makecell}
\usepackage{subcaption}
\usepackage{xstring}
\usepackage{pifont}
\newcommand{\emoji}[1]{%
  \IfStrEq{#1}{prohibited}{%
    \textcolor{black!60}{\ding{55}}%
  }{%
    \IfStrEq{#1}{bust-in-silhouette}{%
      \textcolor{teal}{\textbf{U}}%
    }{%
      \IfStrEq{#1}{robot}{%
        \textcolor{teal}{\textbf{A}}%
      }{%
        \texttt{[#1]}%
      }%
    }%
  }%
}
\usepackage{algorithm}
\usepackage{amsthm}
\usepackage{algorithmic}
\usepackage{dirtree}
\usepackage{enumitem}

\usepackage[acronym,toc,sort=use]{glossaries}
\makeglossaries

\definecolor{teal}{HTML}{1E88E5}   %
\definecolor{coral}{HTML}{D81B60}  %

\usepackage{listings}
\usepackage{mdframed}
\mdfdefinestyle{callout}{%
  leftline=true,
  rightline=false,
  topline=false,
  bottomline=false,
  linewidth=2.5pt,
  linecolor=gray!50,
  backgroundcolor=gray!6,
  innerleftmargin=10pt,
  innerrightmargin=0pt,
  innertopmargin=4pt,
  innerbottommargin=4pt,
  skipabove=3pt,
  skipbelow=3pt,
}

\usepackage[export]{adjustbox}

\title{AC4A: Access Control for Agents}

\author{Reshabh K Sharma}
\affiliation{%
  \institution{University of Washington}
  \city{Seattle}
  \state{Washington}
  \country{USA}}
\email{reshabh@cs.washington.edu}

\author{Dan Grossman}
\affiliation{%
  \institution{University of Washington}
  \city{Seattle}
  \state{Washington}
  \country{USA}}
\email{djg@cs.washington.edu}

\date{March 2026}

\begin{document}

\begin{abstract}
Large Language Model (LLM) agents combine the chat interaction capabilities of LLMs with the power to interact with external tools and APIs. This enables them to perform complex tasks and act autonomously to achieve user goals. However, current agent systems operate on an all-or-nothing basis: an agent either has full access to an API's capabilities and a web page's content, or it has no access at all. This coarse-grained approach forces users to trust agents with more capabilities than they actually need for a given task. For instance, when asking an agent to create an event in a calendar, the user must grant it access to the entire calendar API, which could allow the agent to create events at any time, modify existing events, or access existing events.

In this paper, we introduce \toolname{}, an access control framework for agents. As agents become more capable and autonomous, users need a way to limit what APIs or portions of web pages these agents can access, eliminating the need to trust them with everything an API or web page allows. Our goal with \toolname{} is to provide a framework for defining permissions that lets agents access only the resources they are authorized to access. \toolname{} works across both API-based and browser-based agents. It does not prescribe what permissions should be, but offers a flexible way to define and enforce them, making it practical for real-world systems.

\toolname{} works by creating permissions granting access to resources, drawing inspiration from established access control frameworks like the one for the Unix file system. Applications define their resources as hierarchies and provide a way to compute the necessary permissions at runtime needed for successful resource access. The framework keeps track of currently granted permissions and includes a permission management dashboard for creating and removing them.

We demonstrate the usefulness of \toolname{} in enforcing permissions over real-world APIs and web pages through case studies. The source code of \toolname{} is available at \url{https://github.com/reSHARMA/AC4A}.
\end{abstract}

\maketitle

\section{Introduction}

Large Language Models (LLMs) have evolved from passive question-answering systems to active agents that can interact with the external world through APIs and web browsers~\cite{agentsurvey, webvoyager, osworld}. These agents can now perform complex tasks such as booking travel, managing calendars, or writing code by calling external services and navigating web interfaces~\cite{cursor, operator}. While this autonomy enables powerful applications, it also creates resource misuse challenges that existing systems are ill-equipped to handle.

Unlike traditional software systems, current LLM agents lack an access control framework. While access control systems themselves cannot protect against all security threats, they serve as essential components that enable other security mechanisms. Current agentic systems operate on an all-or-nothing basis: an agent either has full access to an API's capabilities and a web page's content, or it has no access at all. This coarse-grained approach forces users to trust agents with more capabilities than they actually need for a given task. For instance, when asking an agent to create an event in a calendar, the user must grant it access to the entire calendar API, which could allow the agent to create events at any time, modify existing events, or access existing events. Similarly, when using a browser-based agent to check a calendar, the agent gains access to the entire web page, including sensitive information such as other calendar entries and personal details visible to the agent.

The lack of access control is distinct from other security challenges for agents. There are many important security concerns related to agents and the LLMs used underneath, like hallucination~\cite{hallucination-survey}, prompt injection attacks~\cite{long-context-attacks-1, long-context-attacks-2}, and malicious instruction hiding~\cite{rag-attacks-1, rag-attacks-2}. Unlike defenses against these, an access control framework cannot mitigate them directly, similar to how vulnerabilities in traditional systems such as buffer overflows, side-channel attacks, and phishing cannot be addressed by access control systems alone. However, it is unimaginable to think about a secure system without an access control system. We focus on controlling what resources the agent can access in the first place, which we believe is a prerequisite for addressing other security concerns.

We introduce \toolname{}, an access control framework for LLM agents. \toolname{} provides infrastructure for defining fine-grained permissions. Our framework works across both API-based and browser-based agents, using a unified permission infrastructure that can be extended to support new types of agent implementations in the future.

\toolname{} works by creating permissions granting access to resources. Applications define their resources in a hierarchical structure. Permissions in \toolname{} are made up of a resource and application-defined actions such as read, write, or create. The framework includes a permission management dashboard for creating, viewing, and removing permissions. During runtime, when the agent calls an API or interacts with a web page, the call is intercepted and access to the corresponding resources is granted only when there is an explicit permission allowing it.

Our contribution is an access control framework for agents that uses the same permission representation for both API-based and browser-based agents. To the best of our knowledge, \toolname{} is the first framework to provide resource-level access control for browser-based agents, and the first to unify permission enforcement across both agent types. The remainder of this paper is organized as follows. Section~\ref{sec:threat} describes the threat model under which \toolname{} operates. Section~\ref{sec:design} presents the design of \toolname{}, covering permission representation and enforcement. Section~\ref{sec:gen-perm} describes the user interface for managing permissions. Section~\ref{sec:case-study} demonstrates \toolname{} through two case studies: a tic-tac-toe game illustrating core components, and a flight-booking task involving Outlook Calendar, Expedia, and a payment wallet. Section~\ref{sec:limit} discusses limitations and future work. We discuss related work in Section~\ref{sec:related}. Section~\ref{sec:conclusion} concludes.

\section{Threat Model}
\label{sec:threat}

The primary goal of our access control framework is to enforce a given set of permissions for agents. Our threat model is intentionally scoped to the access control layer. We assume that the agent is untrusted, while the access control framework itself and the permissions it receives are trusted. The framework's guarantees are strictly limited to the correct enforcement of the permissions it is given.

\noindent\textbf{Agents and Underlying LLMs.}
The agent is the central untrusted entity. Regardless of its origin or intent, an agent may be compromised or designed with malicious intent. Similarly, the underlying LLMs are not trusted and may contain biases, implementation errors, or backdoors. Even when an agent is well-intentioned, LLMs can simply get things wrong due to a mismatch between the actual user intent and its perception of the intent. For example, if a user asks an agent to \textit{find a good flight}, the agent might actually purchase a flight without explicit authorization. If asked to \textit{find room on my calendar for a meeting}, the agent might cleverly make room by deleting existing events. An access control framework ensures that such unintended actions cannot occur unless the user has explicitly granted the necessary permissions. The value of \toolname{} is ensuring that imperfect instructions or imperfect LLMs do not perform actions the user did not explicitly permit. As an access-control system, \toolname{} cannot ensure that the \textit{correct} action is performed, only that any actions performed are within the allowed permissions.

\noindent\textbf{APIs, Web Pages, and Application Providers.}
We assume that users already trust the applications they use, such as their calendar or a travel-booking website. Our framework controls how agents use these applications, not the applications themselves. We trust applications to accurately map their API endpoints and web page DOM elements to the resources they expose. For example, we trust that a calendar application correctly identifies a call to delete calendar entries on a particular date as requiring write permission for that date. We also trust applications not to leak or corrupt data, just as the Unix file system implementation is trusted in traditional systems.

\noindent\textbf{End Users.}
In this paper, we assume end users grant permissions to agents. Our system would also work in settings where different principals grant permissions. For example, in a corporate setting, permissions might be granted only by supervisors or legal departments. In any case, we trust that the correct permissions are granted. These permissions are handled directly by \toolname{} and agents cannot modify them without end users allowing them to.

Access control is a \textit{necessary but insufficient} tool for achieving higher-level security properties. Our framework's role is not to guarantee that all unintended access to resources is prevented, but rather to enforce the restrictions specified in the provided permission set. If a necessary restriction is omitted or misconfigured by a trusted author, the framework cannot prevent resource misuse. For example, suppose a calendar application includes recurring information in events, such as \textit{weekly}. If an agent has access to the current week but not next week, it might still infer information about events next week from the recurring metadata. Whether this constitutes a flaw in the application's permission function is debatable, but it is beyond the scope of access control enforcement.

\section{Design}
\label{sec:design}

\textit{Permissions} are the core of \toolname{}. A permission in \toolname{} is made up of a \textit{resource value specification} and an \textit{action}. A resource value specification represents a set of \textit{resource values}, and an action represents an operation.
\toolname{} treats resource values as opaque entities without knowing what they represent, but every resource value has a resource type. Applications define resource types using interfaces provided by \toolname{}.

\subsection{Defining Resource Types}
\label{data-rep}
\begin{figure}[ht]
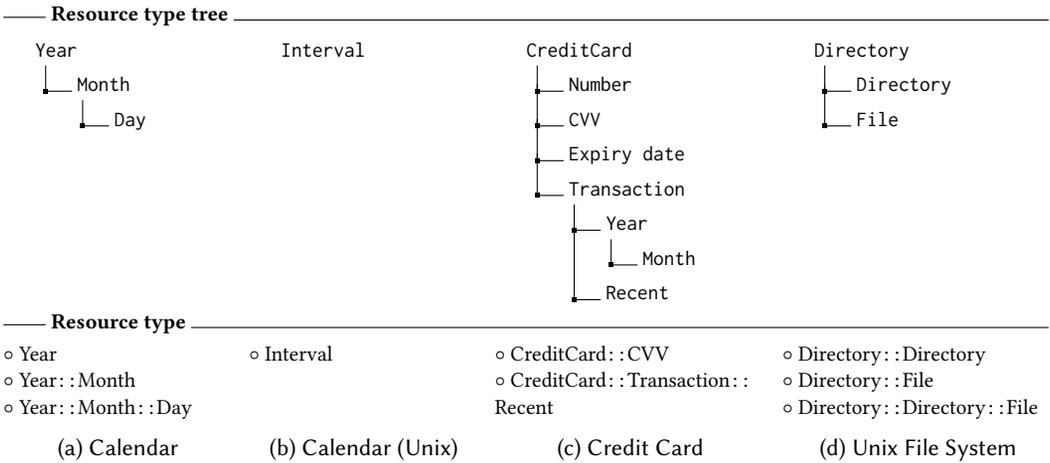

    \centering
    \footnotesize
    \sloppy
    \setlength{\tabcolsep}{3pt}%
    \begin{tabular}[t]{@{} >{\raggedright\arraybackslash}p{0.22\linewidth} >{\raggedright\arraybackslash}p{0.22\linewidth} >{\raggedright\arraybackslash}p{0.26\linewidth} >{\raggedright\arraybackslash}p{0.26\linewidth} @{}}
    \\\noalign{\vspace{-1.5em}}
    \multicolumn{4}{@{}p{\linewidth}@{}}{\rule[0.3ex]{2em}{0.4pt}\ \textbf{Resource type tree}\ \hrulefill}\\[0.25em]
    \begin{minipage}[t]{\linewidth}\dirtree{%
    .1 Year.
    .2 Month.
    .3 Day.
    }\end{minipage}
    &
    \begin{minipage}[t]{\linewidth}\dirtree{%
    .1 Interval.
    }\end{minipage}
    &
    \begin{minipage}[t]{\linewidth}\dirtree{%
    .1 CreditCard.
    .2 Number.
    .2 CVV.
    .2 Expiry date.
    .2 Transaction.
    .3 Year.
    .4 Month.
    .3 Recent.
    }\end{minipage}
    &
    \begin{minipage}[t]{\linewidth}\dirtree{%
    .1 Directory.
    .2 Directory.
    .2 File.
    }\end{minipage}
    \\[0.9em]
    \multicolumn{4}{@{}p{\linewidth}@{}}{\rule[0.3ex]{2em}{0.4pt}\ \textbf{Resource type}\ \hrulefill}\\[0.25em]
    $\circ$ Year\newline $\circ$ Year\texttt{::}Month \newline $\circ$ Year\texttt{::}Month\texttt{::}Day
    &
    $\circ$ Interval
    &
    $\circ$ CreditCard\texttt{::}CVV \newline $\circ$ CreditCard\texttt{::}Transaction\texttt{::}\newline Recent
    &
    $\circ$ Directory\texttt{::}Directory\newline $\circ$ Directory\texttt{::}File\newline $\circ$ Directory\texttt{::}Directory\texttt{::}File
    \\\noalign{\vspace{-1.0em}}
    \subcaption{Calendar}
    &
    \subcaption{Calendar (Unix)}
    &
    \subcaption{Credit Card}
    &
    \subcaption{Unix File System}
    \end{tabular}
    \vspace{-2.2em}
    \caption{Resource type trees and derived resource types. (a) and (b) show two alternative representations for calendar resources. (c) shows a credit card model with overlapping access paths. (d) shows a recursive file system model.}
    \label{fig:resource-examples}
\end{figure}

A resource type is defined as a directed path originating from the root of a resource type tree. It represents a set of resource values and is made up of \textit{resource type nodes} chained together using the \texttt{::} operator. Edges in a resource type tree represent parent-child relationships such that the set of resource values represented by the parent is a superset of the resource values represented by the child. In Figure~\ref{fig:resource-examples}, we show resource type trees and examples of resource types for calendar, file system, and credit card information. Applications can model multiple resource type trees for decomposing the same resources in different ways, as shown in Figure~\ref{fig:resource-examples}(a) and (b), where calendar events are represented by date in (a) or by Unix timestamp intervals in (b).

Using Figure~\ref{fig:resource-examples}(a), the three resource types are Year, Year::Month, and Year\texttt{::}Month\texttt{::}Day. We expect (but \toolname{} does not enforce) that the different resource values of a particular type such as Year::Month are disjoint since the resources in March, 2026 are disjoint from the resources in February, 2026. The tree structure does imply and assume the resources in 2026 are a superset of both March, 2026 and February, 2026. Disjoint resource values are not required nor always desired. For example, the Interval resource type in Figure~\ref{fig:resource-examples}(b) has values that can represent any time interval (and thus a set of resource values can represent any set of intervals). It is also not required for resource types to be disjoint and they can also have common resource values. The same resource type tree can mix disjoint and overlapping approaches. For example, in Figure~\ref{fig:resource-examples}(c) we expect the CVV, Expiry date, and Transaction components of a credit card are disjoint resources, but within transactions, permission can be granted either for a particular year or month or for the k most Recent transactions, which are two different ways of providing access to only some of the transaction history.

Multiple resource type nodes in a resource type can have the same name, making a resource type tree recursive and resulting in an infinite tree. For example, in Figure~\ref{fig:resource-examples}(c) the \textit{Directory} node can be present multiple times. Even in this case when a resource type tree is infinite, resource types are still finite as they are fixed-length paths from the root.

The primary purpose of resource types is to represent parent-child relationships, not to enforce semantic validity. \toolname{} neither defines nor validates legal values for any resource type. For example, a type \texttt{Year\texttt{::}Month} can represent a value \textit{3}, \textit{March}, or even \textit{Marchuary}. Similarly, \toolname{} treats an \texttt{Interval} representing Unix timestamps as an opaque string. Applications may enforce their own constraints on these values as needed. The reason we do not need a system for semantic validity is that our goal is simply to check, as we describe in Section~\ref{ssec:perm-check}, whether permissions granted to an agent suffice for the permissions required by an action. Granting permission to Year(-3)\texttt{::}Month(Marchuary) is harmless unless some API call requires such a permission to proceed.

\subsection{Resource Values \& Resource Value Specifications}
\label{sec:rvs}
A \textit{resource value specification} represents a set of resource values. It is written as TypeNode1(\textit{Val1}) \texttt{::}TypeNode2(\textit{Val2})\texttt{::}\ldots\texttt{::}TypeNodeN(\textit{ValN}), where each \textit{Val} is either a string or the wildcard \texttt{?} (denoting the union of all possible values for that resource type node), and each TypeNode is a resource type node chained with the \texttt{::} operator along a path in a resource type tree. In Figure~\ref{fig:resource-value-sets}, we show examples of resource value sets represented by corresponding resource value specifications.

\begin{figure}[t]
    \centering
    \fontsize{8.5}{10}\selectfont
    \begin{tabular}{@{}>{\raggedright\arraybackslash}p{0.48\textwidth}>{\raggedright\arraybackslash}p{0.48\textwidth}@{}}
    \toprule[1.5pt]
    \textbf{Resource value specification} & \textbf{Set of resource values} \\
    \midrule[1.5pt]
    Year(2026)\texttt{::}\allowbreak Month(October)\texttt{::}\allowbreak Day(15) &
    Calendar events on 15th October 2026 \\
    \midrule[0.5pt]
    Year(\texttt{?})\texttt{::}\allowbreak Month(January) &
    Calendar events in January across all years \\
    \midrule[0.5pt]
    Year(2026)\texttt{::}\allowbreak Month(\texttt{?})\texttt{::}\allowbreak Day(15) &
    Calendar events on the 15th of all months in 2026 \\
    \midrule[0.5pt]
    Interval(1696809600\allowbreak--1696896000) &
    Calendar events between 9th and 10th October 2023 \\
    \midrule[0.5pt]
    Directory({\hypersetup{urlcolor=black}\url{/}})\texttt{::}Directory(home)\texttt{::}\allowbreak Directory(user)\texttt{::}\allowbreak File(report.txt) &
    File at path {\hypersetup{urlcolor=black}\url{/home/user/report.txt}} \\
    \midrule[0.5pt]
    Directory({\hypersetup{urlcolor=black}\url{/}})\texttt{::}Directory(home)\texttt{::}\allowbreak Directory(\texttt{?})\texttt{::}\allowbreak File(report.txt) &
    All files named report.txt in any immediate subdirectory of home \\
    \midrule[0.5pt]
    CreditCard(\texttt{?})\texttt{::}\allowbreak Transaction(\texttt{?})\texttt{::}\allowbreak Recent(5) &
    The last 5 transactions across all credit cards \\
    \midrule[0.5pt]
    CreditCard(``GoldPlus'')\texttt{::}\allowbreak Transaction(\texttt{?})\texttt{::}\allowbreak Year(2026) \texttt{::} Month(January) &
    All transactions in January 2026 on the GoldPlus credit card \\
    \bottomrule[1.5pt]
    \end{tabular}
    \caption{Resource value specifications and the corresponding set of resource values.}
    \label{fig:resource-value-sets}
\end{figure}

\subsection{Defining Actions}
\label{sec:action}
\toolname{} uses a simple access control model where each application defines its own set of actions. An \textit{action} represents a type of operation that can be performed on resources. Unlike resource type nodes, actions have no structure and are assumed to be non-overlapping. Most applications will likely have \textit{read}, \textit{write}, and \textit{create} actions, but applications are free to define additional actions as needed. The use of read/write/create as actions is merely conventional and \toolname{} does not enforce any meaning on actions. Even though the read/write/create convention is admittedly simple and not particularly precise, it works adequately for most applications. \textit{Read} covers any operation that makes resources visible to an agent without modifying the underlying resource store, including resource retrieval and querying operations, search, resource display, and read-only API calls. \textit{Write} access covers operations that modify the internal state of resource systems, including resource modification and deletion. \textit{Create} allows creating a new resource instance without issuing a blanket write access for that resource. For example, a calendar application might define \texttt{read} to allow viewing calendar events, \texttt{write} to allow modifying or deleting events, and \texttt{create} to allow only creating new events. A file system application might use \texttt{\{read, write\}}, where \texttt{read} covers file retrieval and directory listing, and \texttt{write} covers file creation, modification, and deletion. Applications are free to define additional actions like \texttt{delete} and \texttt{modify}.

Permissions in \toolname{} are a pair of a resource value specification (Section~\ref{sec:rvs}) and an action (Section~\ref{sec:action}).
When a resource value specification contains \texttt{?}, that permission represents the union of all permissions where \texttt{?} is instantiated with each possible value for that resource type node.
In Figure~\ref{fig:permission-examples}, we show examples of permissions in \toolname{}.

\begin{figure}[h]
    \centering
    \fontsize{8.5}{10}\selectfont
    \begin{tabular}[t]{p{0.29\textwidth}>{\centering\arraybackslash}p{0.11\textwidth}>{\raggedright\arraybackslash}p{0.52\textwidth}}
    \toprule[1.0pt]
    \textbf{Resource Value Specification} & \textbf{Action} & \textbf{Description} \\
    \midrule[1.0pt]
    Year(2026) & \texttt{write} & Access for writing to resources of type Year with value 2026. \\
    \midrule[0.5pt]
    Year(?)\texttt{::}Month(January)  & \texttt{read} & Access for reading all resources of type Month with value January across all years. \\
    \midrule[0.5pt]
    \makecell[tl]{Directory(home)\texttt{::}Directory(?)\\\texttt{::}File(report.txt)} & \texttt{read} & Access for reading all files named report.txt in any immediate subdirectory of home in a file system model. \\
    \midrule[0.5pt]
    \makecell[tl]{CreditCard(?)\texttt{::}Transaction(?)\\\texttt{::}Recent(5)} & \texttt{read} & Access for reading the last 5 transactions across all credit cards in the credit card model. \\
    \bottomrule[1.0pt]
    \end{tabular}
    \caption{Example permissions. Each permission is a pair of a resource value specification and an action.}
    \label{fig:permission-examples}
\end{figure}

\subsection{Permission Checking}\label{ssec:perm-check}
When an agent attempts to access a resource, \toolname{} must determine whether the agent's currently granted permissions are sufficient. This requires knowing two things: (1) what permissions are necessary for the access, and (2) what permissions have been granted to the agent. Section~\ref{sec:dyn-perm-sig} explains how \toolname{} determines the necessary permissions for an agent to proceed. Section~\ref{sec:gen-perm} describes the user interface through which users grant permissions. In this section, we assume both are already known and explain how to check if the granted permissions are sufficient to allow the access.

We focus on whether the set of resource values represented by the granted resource value specifications \textit{covers} the set of resource values needed by the agent for a specific action. This check is repeated separately for each action to ensure that permissions for one action cannot satisfy the needs of another. To determine if the granted set covers the needed set, \toolname{} uses a resource\_difference function defined by each application. \toolname{} uses resource\_difference as it iterates through the granted permissions, tracking as it iterates what portions of the necessary permissions are not yet known to be granted.

\subsubsection{Resource Difference Function}
\label{sec:residual}
Each application implements a resource\_difference function with the following signature:
\begin{align*}
\mathrm{resource\_difference}(\textit{Need},\,\textit{Have}) \Rightarrow \textit{Remaining}
\end{align*}
where \textit{Need} and \textit{Have} are resource value specifications (which may contain \texttt{?}), and \textit{Remaining} is a finite set of resource value specifications whose union denotes what part of \textit{Need} is still unmet after accounting for \textit{Have}. When either argument contains \texttt{?}, the function must reason about the union of all possible instantiations represented by that specification.

\toolname{} performs permission checking using the resource\_difference function for each action from the need set. It maintains a working set of unmet resource value specification requirements. Starting with the needed resource value specification, we iteratively subtract what each granted resource value specification covers until either nothing remains (need met) or we exhaust all granted resource value specifications (need unmet). Access is allowed only when the need becomes empty at the end for each needed action. Algorithm~\ref{alg:perm-check} formalizes this approach, where $getResources$ returns a list of resource value specifications associated with a specific action, $Need$ is the set of required resource value specifications for an action $Action$ from the set of needed actions $NeedActions$, and $HaveResources$ is the set of granted resource value specifications from the active permissions with action $Action$.
\begin{algorithm}[h]
\caption{\toolname{}'s Permission Checking Algorithm}
\label{alg:perm-check}
\begin{algorithmic}[1]
\FOR{$Action \in NeedActions$}
    \STATE $HaveResources \gets \mathrm{getResources}(ActivePermissions, Action)$
    \STATE $Remaining \gets \{Need\}$
    \FOR{each $have$ in $HaveResources$}
        \STATE $Next \gets \emptyset$
        \FOR{each $n$ in $Remaining$}
            \STATE $Next \gets Next \cup \mathrm{resource\_difference}(n, have)$
        \ENDFOR
        \STATE $Remaining \gets Next$
        \IF{$Remaining = \emptyset$}
            \STATE \textbf{break}
        \ENDIF
    \ENDFOR
    \IF{$Remaining \neq \emptyset$}
        \STATE \textbf{return} \textbf{false}
    \ENDIF
\ENDFOR
\STATE \textbf{return} \textbf{true}
\end{algorithmic}
\end{algorithm}

Algorithm~\ref{alg:perm-check} assumes that the resource\_difference function provided by the applications is sound. This means it does not allow access to unauthorized resources. To achieve this, the function need not precisely return the exact remaining resources after accounting for granted resources. As long as it never returns resources outside of what was originally needed, the algorithm remains sound. For example, if an agent needs access to \{\textit{tag1}, \textit{tag2}, \textit{tag3}\} and has resource \{\textit{tag1}\} for the same action, the resource\_difference function could return either \{\textit{tag2}, \textit{tag3}\} (precise) or \{\textit{tag1}, \textit{tag2}, \textit{tag3}\} (conservative). Both implementations are sound since they do not grant access to anything beyond what was needed.

\textit{Order independence} is a property of the resource\_difference function ensuring that the final permission checking decision remains the same regardless of the order in which granted resources are processed. \toolname{} ensures that the permission checking is always sound if the resource\_difference function never returns resources outside of the originally needed resources, but this does not guarantee order independence. Applications whose resource\_difference function does not return the precise set of remaining resources must ensure that the function is order independent. For example, if the resource\_difference function precisely computes interval subtraction for the calendar interval resource type node, it is order independent. If the function sometimes over-approximates coverage (e.g., by rounding up intervals), different orderings of granted resources can produce different final \textit{Remaining} sets, breaking order independence, though the algorithm remains sound.

Application developers should ensure that their resource\_difference implementation behaves consistently regardless of the order in which granted resources are processed in Algorithm~\ref{alg:perm-check}. Set-based resources naturally achieve this property. Hierarchical and interval-based resources require consistent refinement or conservative handling. Applications can choose their precision-performance trade-off but should ensure their choice does not introduce order dependence and remains sound. \toolname{} provides support for common patterns like conservatively finding the remaining resources for enums, intervals, and trees, making the implementation of the resource\_difference function easier. These helper functions do not interpret the actual data represented by the resource types and only handle simple cases, such as when the need is a subtree of \textit{have} or when the enum values of \textit{need} and \textit{have} are the same. Applications are free to implement their own resource\_difference functions for more complex scenarios like overlapping resource types or non-hierarchical relationships. For example, in Figure~\ref{fig:resource-examples}(c), the same credit card transaction can be accessed via recency-based and time-based paths. The resource\_difference function can handle this overlap in different ways:

\begin{itemize}
    \item \textbf{Conservative approach}: The function might treat the recency and time-based dimensions as incomparable. Having access via the recency-based resource would not be considered to cover the need for the time-based resource and resource\_difference might conservatively return the full need for the time-based resource.
    \item \textbf{Precise approach}: The function might recognize that time-based access covers all transactions in that month. When an agent has both recency-based and time-based access that overlap, the function could compute that the need is satisfied for the overlapping data and return the empty set.
\end{itemize}

Both approaches are valid, and the choice is entirely up to the application. The framework only requires that the resource\_difference function is sound and does not violate order independence.

\subsection{Sufficient Permissions for an Access}
\label{sec:dyn-perm-sig}

When an agent performs an access, such as calling an API endpoint or interacting with a web page element, we need to know what permissions are sufficient for that access. Given that information, we use the resource\_difference function (Section~\ref{sec:residual}) to compute whether the currently granted permissions are enough. Sufficient permissions are determined differently for APIs and web pages. We first explain the process for APIs and then for web pages.

\subsubsection{APIs}
\label{sec:perm-func}

For API-based agents, each application provides a \textit{permission function} for its API endpoints. The permission function takes an endpoint name and the actual argument values of a call, and returns a finite set of permissions (resource value specification and action pairs) that are sufficient for that call. Figure~\ref{fig:signature-examples} shows examples of permissions returned by permission functions for different API calls.

\begin{figure}[h]
    \centering
    \small
    \sloppy
    \begin{tabular}{>{\raggedright\arraybackslash}p{0.57\textwidth}>{\raggedright\arraybackslash}p{0.38\textwidth}}
    \toprule[1.0pt]
    \textbf{API Call} & \textbf{Sufficient Permissions} \\
    \midrule[1.0pt]
    get\_calendar\_events(start\_date: 2024-01-01, end\_date: 2024-01-31) & Year(2024)\texttt{::}\allowbreak Month(January), \texttt{read} \\
    \midrule[0.5pt]
    get\_events(start\_time: 1696809600, end\_time: 1696896000) & Interval(1696809600-1696896000), \texttt{read} \\
    \midrule[0.5pt]
    read\_file(path: "/home/user/docs/\allowbreak report.txt") & Directory(home)\texttt{::}\allowbreak Directory(user)\texttt{::}\allowbreak Directory(docs)\texttt{::}\allowbreak File(report.txt), \texttt{read} \\
    \bottomrule[1.0pt]
    \end{tabular}
    \caption{Examples of sufficient permissions returned by permission functions for API calls.}
    \label{fig:signature-examples}
\end{figure}

\toolname{} provides the enforcement infrastructure but has no knowledge of an application's API semantics or what resources a given endpoint actually accesses. The permission function may be as precise or as approximate as the application developer chooses, and \toolname{} is entirely agnostic to this choice. For example, if an API call accesses calendar events from January~1 to February~15, 2026, the permission function could return day-level resource value specifications covering exactly those 46~days, or it could round up to month-level granularity and return \{Year(2026)\texttt{::}Month(January), Year(2026)\texttt{::}Month(February)\}.

Correctness of the permission function is entirely the responsibility of the application. \toolname{} trusts the permission function's output and has no independent way to verify that it accurately reflects the resources an endpoint actually accesses. If a calendar permission function reports that an endpoint only needs January when it in fact reads January through mid-February, \toolname{} will enforce the narrower (incorrect) requirement, and the endpoint may access resources beyond what the user intended to authorize. The permission function must also handle unmapped endpoints. A conservative choice is to require broad permissions so that unmapped calls are blocked unless the agent has been granted wide access. Figure~\ref{fig:api-mapping} shows an example permission function for a calendar application.

\begin{figure}[ht]
    \centering
    \begin{minted}[fontsize=\footnotesize, frame=single, breaklines=true]{python}
def permission_function(kwargs, endpoint_name):
    resources = []
    start_time = kwargs['start_time']
    duration = kwargs['duration']
    end_time = start_time + duration
    month_names = ['January', 'February', 'March', 'April', 'May', 'June',
                    'July', 'August', 'September', 'October', 'November', 'December']
    def get_hierarchy(time_value):
        return [
            (365, 'Year', time_value.year),
            (calendar.monthrange(time_value.year, time_value.month)[1], 'Month',
                month_names[time_value.month - 1]),
            (1, 'Day', time_value.day),
            (0, 'Hour', time_value.hour)
        ]
    for (_, label, start_value), (_, _, end_value) in zip(
        get_hierarchy(start_time), get_hierarchy(end_time)):
        resources.append(f'Calendar:{label}({start_value})')
        if start_value != end_value:
            break
    action = (
        "create"
        if any(k in endpoint_name for k in ["reserve", "create", "add"])
        else "write"
        if any(k in endpoint_name for k in ["update", "edit", "modify"])
        else "read"
    )
    return '::'.join(resources), action
    \end{minted}
    \caption{Example permission function for a calendar application. The function takes an endpoint name and argument values, and returns resource value specifications and an action.}
    \label{fig:api-mapping}
\end{figure}

As applications evolve, resource type trees can be extended without breaking existing permissions. For example, if a calendar application adds functionality for managing working hours, a new \textit{WorkingHours} node can be added as a child of \textit{Day} and the permission function updated to return it for relevant endpoints. Existing permissions that grant access to calendar events continue to work. A permission covering a \textit{Day} node implicitly covers the new \textit{WorkingHours} child, so existing broad permissions remain valid without modification.

\subsubsection{Web Pages}

For browser-based agents, we ensure that the agent can view and interact with some DOM elements but not others. In principle, we could treat every DOM element as an API entry point and have a permission function that returns the permissions needed to access it. However, web pages often contain thousands of DOM elements, so specifying permissions for each separately is impractical.

Instead, we map CSS-like selectors to resource value specifications and actions separately. A configuration file for each web page contains entries that associate selectors with resource value specifications and actions. For a given DOM element, the required permissions are computed for any selector that matches that element. Elements that match no selector require no permission and remain visible by default. This is practical because most web page elements (headings, logos, navigation) do not contain protected resources.

Whenever the DOM changes (due to navigation, JavaScript execution, or agent actions) or the agent's granted permissions change, \toolname{} re-evaluates the configuration against the current DOM and active permissions. Elements whose required permissions are not covered by the agent's active permissions are hidden before the agent sees the page.

In its simplest form, a configuration file maps plain CSS selectors to resource value specifications or actions. For example, a calendar page might map the calendar grid showing events from June 2026, \texttt{\#calendarGrid} to Year(2026)\texttt{::}Month(June) and also to read action. Similarly, it might map a button for saving an event, \texttt{\#saveEventBtn}, to Year(?) and also to the create action.

Real web pages introduce two complications that require going beyond plain CSS selectors. First, many sites dynamically generate HTML using JavaScript and randomize IDs and class names to protect against data scraping, making simple selectors like \texttt{\#calendarGrid} unreliable across sessions. To handle this, configuration files use extended CSS selectors that support queries based on stable element attributes and content, such as \texttt{button[aria-label*='select to change the month']} or \texttt{span:contains('New event')}.

Second, a single HTML element may represent different resource values depending on dynamic page content. A calendar might display different months using the same element structure with different data, or a file system browser might list different directories using the same template. For these cases, the resource value specification itself must be computed from page content at runtime using \textit{resource value string specifications}. Figure~\ref{fig:web-mapping} shows a complete configuration for Outlook Calendar that demonstrates both extended selectors and dynamic resource value string specifications.

\begin{figure}[h]
    \centering
    \begin{minted}[fontsize=\footnotesize, frame=single, breaklines=true]{json}
{
    "https://outlook.live.com/calendar/0/view/workweek": {
        "verified": true,
        "read": [
          "div.ZlutZ"
        ],
        "write": [
          "span:contains('Save')"
        ],
        "create": [
          "span:contains('New event')"
        ],
        "data": {
          "Year($data{button[aria-label*='select to change the month'] > span}[1])::
          Month($data{button[aria-label*='select to change the month'] > span}[0])::
          Day(?)": [
                "div.ZlutZ"
          ],
          "Year($data{input[aria-label='Start date']}{split_slash}[2]@value)::
          Month($data{input[aria-label='Start date']}{split_slash}[0](number_to_month)@value)::
          Day($data{input[aria-label='Start date']}{split_slash}[1]@value)": [
            "span:contains('Save')"
          ],
          "Year(?)": [
            "span:contains('New event')"
          ]
        }
    }
}

    \end{minted}
    \caption{Example web mapping configuration for Outlook Calendar. Static mappings associate CSS selectors with actions, while dynamic mappings use the \$data syntax to extract resource values from page content at runtime.}
    \label{fig:web-mapping}
\end{figure}

\paragraph{Resource Value String Specification}
A resource value string specification extracts a resource value from page content at runtime. The syntax is
\begin{align*}
\$data\{\text{css selector}\}\{\text{list transformation}\}[\text{Index}]\{\text{value transformation}\}@\text{attr}
\end{align*}

This syntax extracts data from the specified CSS selector and attribute (such as \texttt{value}, \texttt{href}, \texttt{src}, \texttt{alt}, \texttt{title}, or \texttt{text}). The data is transformed by the list transformation into a list of strings, then the index selects one string, which can be further transformed by an optional value transformation (written in parentheses after the index, e.g., \texttt{(number\_to\_month)}). Multiple transformations can be chained.

For example:
\begin{align*}
\text{Year}(\$data\{\text{input[aria-label='Start date']}\}\{\texttt{split\_slash}\}[2]@\text{value})
\end{align*}
This selector points to an input element with a value like MM/DD/YYYY. The value is split on the \texttt{/} character, and the element at index 2 (the year) becomes the value for the Year resource type node.

Alternatively, web developers can annotate their pages directly using HTML attributes: \texttt{data-ac4a-resource} for the resource value specification, \texttt{data-ac4a-action} for the action, and \texttt{data-ac4a-static} to indicate whether the resource value is dynamic. When \texttt{data-ac4a-static} is false, the value in a \texttt{data-ac4a-resource} is expected to be a resource value string specification and is evaluated at runtime.

\section{User Interface for Granted Permissions}
\label{sec:permission_ui}
\label{sec:gen-perm}

\begin{figure}[h]
    \centering
    \includegraphics[width=\linewidth]{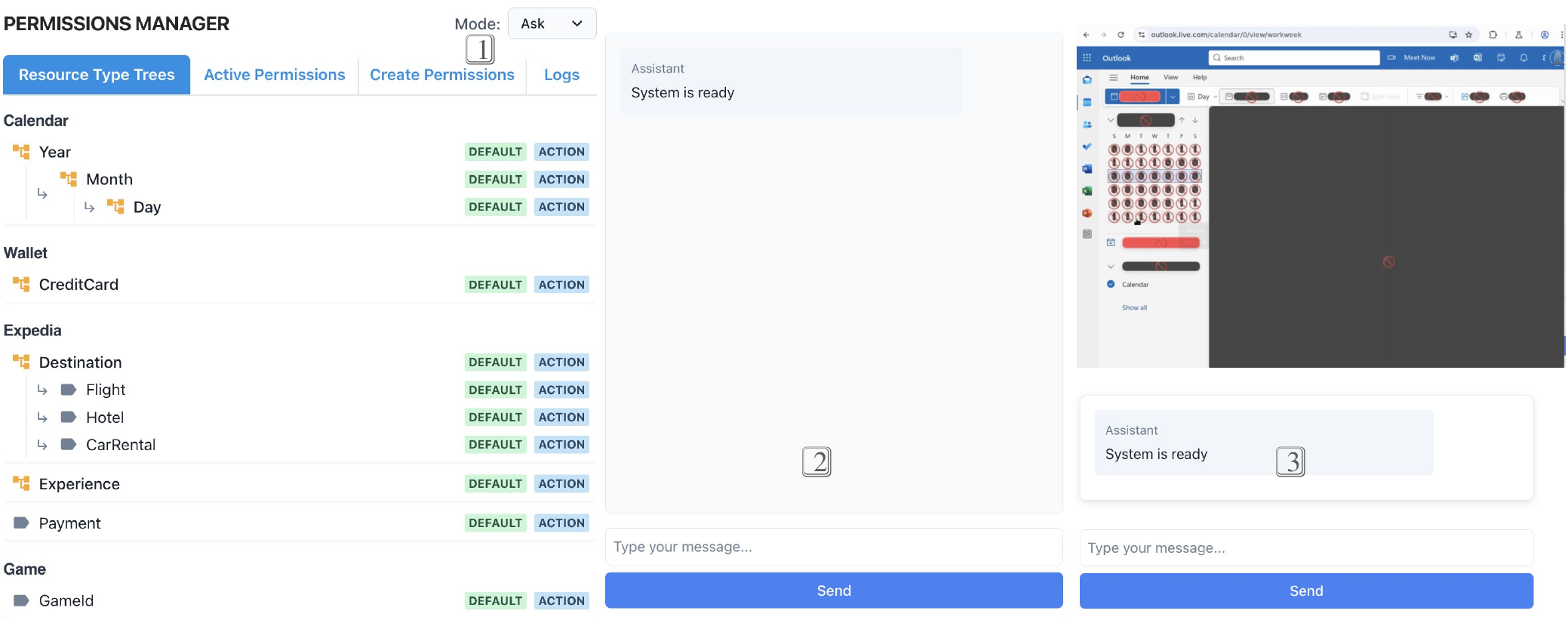}
    \caption{\toolname{} interface. The dashboard (left) provides tabs for browsing resource type trees, viewing active permissions, creating permissions, and reviewing logs. (1)~permission handling mode selector, (2)~chat interface for the API-based agent, (3)~browser preview for the browser-based agent.}
    \label{fig:prototype-interfaces}
\end{figure}

We have implemented a prototype of \toolname{} as shown in Figure~\ref{fig:prototype-interfaces}. The system is built around a chatbot backed by an Autogen-based multi-agent system~\cite{autogen} with access to application endpoints such as a calendar and a travel service. These endpoints are backed either by LLM-based simulated responses or by real API endpoints. Each application provides its resource types, actions, permission function, and resource\_difference function. The same set of active permissions is enforced simultaneously for both the API-based and browser-based agent.

The prototype interface includes a permission manager, a chat interface for interacting with the API-based agent, a live browser view showing what the browser-based agent sees, and optionally the full webpage view (visible to the end user but not the agent). For both agent types, we use GPT-5.4-mini as the underlying model. For the browser agent, GPT-5.4-mini generates browser interactions as python code using pyautogui~\cite{pyautogui}.

\begin{figure}[h]
    \centering
    \begin{minipage}[t]{0.48\textwidth}
        \centering
        \includegraphics[width=0.95\linewidth]{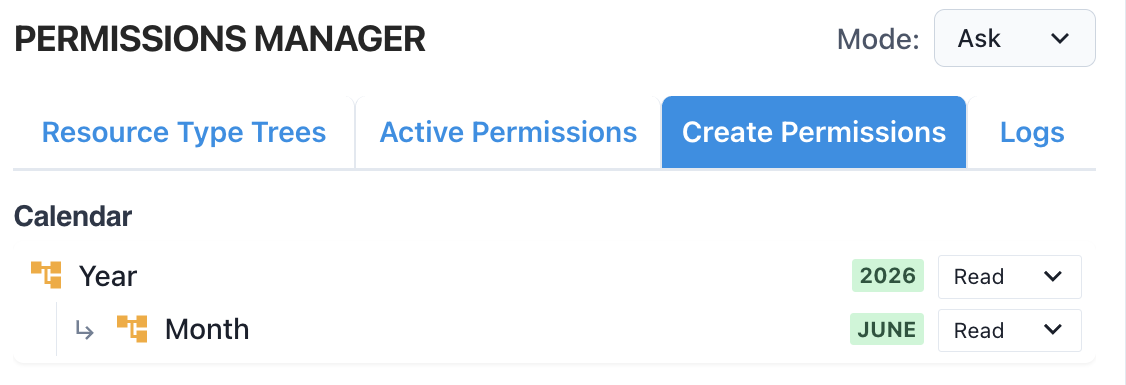}
        \caption{Manual permission creation. Users select resource type nodes, values, and actions from the dashboard.}
        \label{fig:create-perm-manual}
    \end{minipage}
    \hfill
    \begin{minipage}[t]{0.48\textwidth}
        \centering
        \includegraphics[width=0.95\linewidth]{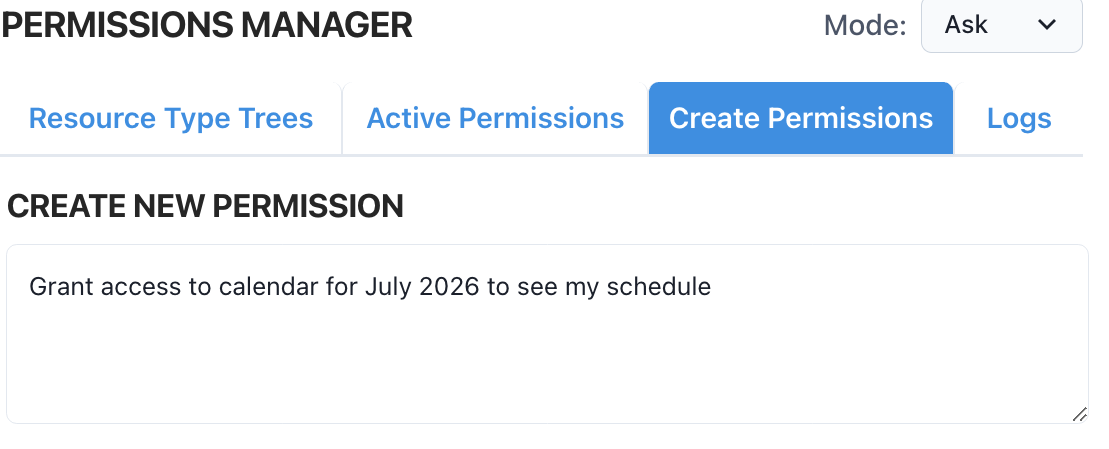}
        \caption{LLM-assisted permission generation. A natural language description is converted to a formal permission.}
        \label{fig:create-perm-llm}
    \end{minipage}
\end{figure}

\begin{figure}[h]
    \centering
    \begin{minipage}[t]{0.48\textwidth}
        \centering
        \includegraphics[width=0.95\linewidth]{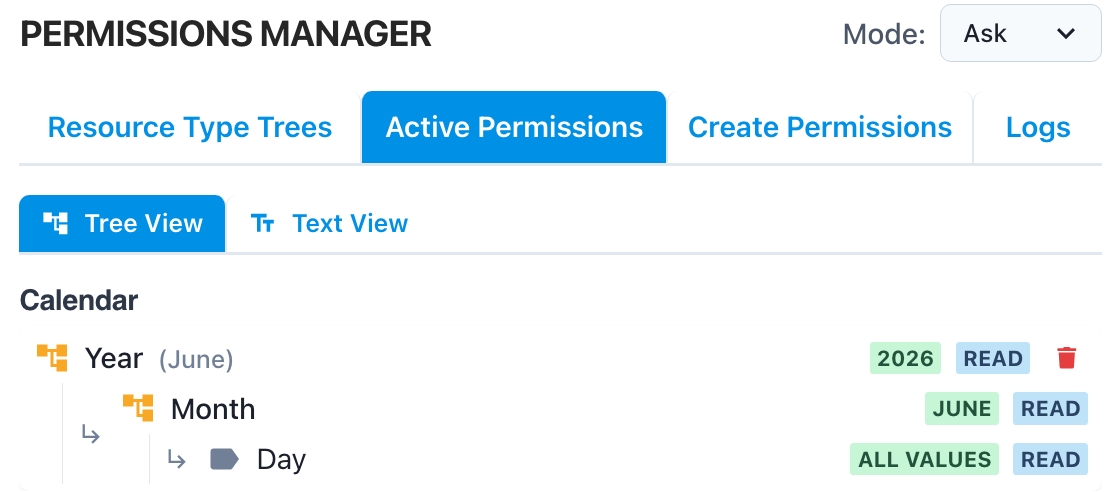}
        \caption{Active permissions displayed in a structured, hierarchical interface for easy review.}
        \label{fig:active-perm}
    \end{minipage}
    \hfill
    \begin{minipage}[t]{0.48\textwidth}
        \centering
        \includegraphics[width=0.95\linewidth]{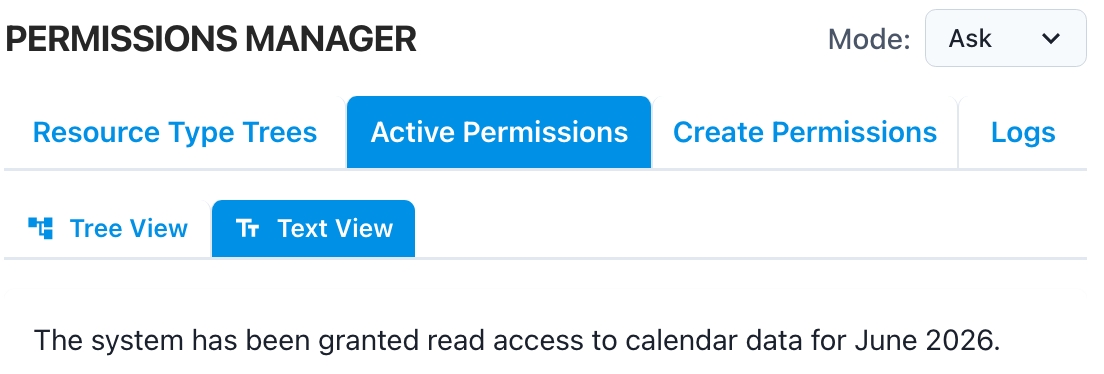}
        \caption{Active permissions rendered as natural language statements.}
        \label{fig:active-perm-nl}
    \end{minipage}
\end{figure}

\toolname{} provides a dashboard for creating, viewing, and removing active permissions. Users can create permissions manually through the dashboard (Figure~\ref{fig:create-perm-manual}) by selecting resource type nodes from the application's resource type trees, providing the corresponding resource value specifications, and choosing an action. Created permissions appear in the active permissions view (Figure~\ref{fig:active-perm}), which displays them in a structured hierarchy organized by resource value specifications and actions. Users can remove permissions from this view at any time. A log view records the complete audit trail of permission creation, removal, and usage, including which permissions allowed or blocked each access. When an access is denied, the dashboard displays the \textit{Remaining} need produced by the resource\_difference function (Section~\ref{sec:residual}), guiding the user toward what permissions to add for the access to succeed on retry.

Beyond manual creation of permissions, \toolname{} supports two additional modalities. LLM-assisted generation (Figure~\ref{fig:create-perm-llm}) converts natural language descriptions into formal permissions. For example, a user can write \textit{grant access to calendar for July to see my schedule}, and the system generates the corresponding permission against the application's resource type trees. Users review and optionally modify the generated permission before it is added. Permissions can also be inferred from user requests or execution plans: the system analyzes implicit access requirements in task descriptions and feeds the inferred needs through the same LLM-assisted pipeline. Active permissions can additionally be rendered as natural language statements (Figure~\ref{fig:active-perm-nl}) for a more human-readable overview.


The system also uses the resource\_difference function to detect redundancy when adding permissions. For example, in a calendar application, if read access to Year(2026) already exists, adding Year(2026)\texttt{::}Month(October) is flagged as redundant because permission to a year implicitly grants access to all its months. However, if the October permission is created first, both permissions coexist. Removing one does not guarantee that the access it covered is revoked, since the other permission may still cover it. The user is responsible for managing such overlaps.

\subsection{Permission Handling Modes}
\label{sec:handling-modes}
Access denials are expected during normal operation. An agent may attempt actions beyond its current authorization, either because the user has not yet granted the necessary permissions or because the agent is exploring available options. Rather than treating denials as errors, \toolname{} provides configurable modes for how the agent should respond.

When the permission checking algorithm determines that the agent's active permissions are insufficient and denies an access, the agent must be made aware so it can respond appropriately. For browser-based agents, blocked areas marked with a \emoji{prohibited} sign signal that the agent lacks the necessary permissions. For API-based agents, the system returns an error message describing the denied access. \toolname{} supports four modes for how the agent should handle denied access, configured through the agent's system prompt.

\begin{enumerate}
    \item \textbf{Ask:} The agent requests the user to grant the necessary permissions and, if available, shares the specific access that was denied so the user can configure permissions themselves.
    \item \textbf{Skip:} The agent suggests a workaround that avoids the unauthorized resource or asks the user how to proceed.
    \item \textbf{Infer:} The system automatically infers the needed permissions using an LLM but does not deploy them without user approval. The agent presents the inferred permissions and waits for confirmation.
    \item \textbf{YOLO:} The system automatically infers and deploys permissions using an LLM. The agent retries the denied action until the correct permissions are set automatically.
\end{enumerate}

Regardless of the mode, all generated and deployed permissions are logged along with their usage to allow or deny access. Users can monitor which permissions are being created and used at any time, providing oversight even when an LLM generates permissions automatically.

These modes represent a trade-off between security and convenience. Manual permission creation combined with Ask or Skip mode is the most secure configuration, as the user retains full control over what permissions are granted. The LLM-assisted modes (Infer and YOLO) provide convenience but introduce potential attack vectors, since the LLM that generates permissions could be manipulated through prompt injection~\cite{prompt-injection, prompt-injection-2} or other techniques. In fact, YOLO is not secure at all beyond logging that the agent added permissions. It is appropriate only when auditing suffices or for experimentation to see how often the agent could have proceeded without user intervention.

\section{Case Studies}
\label{sec:case-study}

\begin{figure}[t]
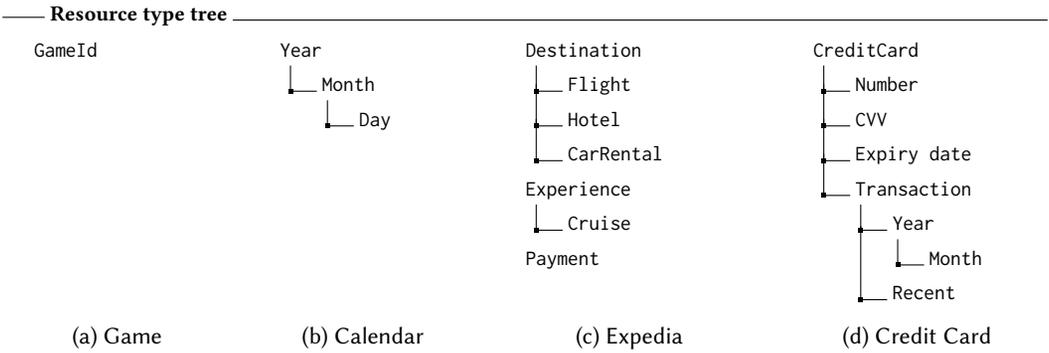

    \centering
    \footnotesize
    \sloppy
    \setlength{\tabcolsep}{3pt}%
    \begin{tabular}[t]{@{} >{\raggedright\arraybackslash}p{0.22\linewidth} >{\raggedright\arraybackslash}p{0.22\linewidth} >{\raggedright\arraybackslash}p{0.26\linewidth} >{\raggedright\arraybackslash}p{0.26\linewidth} @{}}
    \\\noalign{\vspace{-1.5em}}
    \multicolumn{4}{@{}p{\linewidth}@{}}{\rule[0.3ex]{2em}{0.4pt}\ \textbf{Resource type tree}\ \hrulefill}\\[0.25em]
    \begin{minipage}[t]{\linewidth}\dirtree{%
    .1 GameId.
    }\end{minipage}
    &
    \begin{minipage}[t]{\linewidth}\dirtree{%
    .1 Year.
    .2 Month.
    .3 Day.
    }\end{minipage}
    &
    \begin{minipage}[t]{\linewidth}\dirtree{%
    .1 Destination.
    .2 Flight.
    .2 Hotel.
    .2 CarRental.
    }
    \dirtree{%
    .1 Experience.
    .2 Cruise.
    }
    \dirtree{%
    .1 Payment.
    }
    \end{minipage}
    &
    \begin{minipage}[t]{\linewidth}\dirtree{%
    .1 CreditCard.
    .2 Number.
    .2 CVV.
    .2 Expiry date.
    .2 Transaction.
    .3 Year.
    .4 Month.
    .3 Recent.
    }\end{minipage}
    \\\noalign{\vspace{-1.0em}}
    \subcaption{Game}
    &
    \subcaption{Calendar}
    &
    \subcaption{Expedia}
    &
    \subcaption{Credit Card}
    \end{tabular}
    \vspace{-1.5em}
    \caption{Resource type trees for the case study applications. (a)~The tic-tac-toe game uses a single GameId node. (b)~Calendar with Year, Month, and Day resource type nodes. (c)~Expedia models travel with three resource type trees for destinations, experiences, and payment. (d)~Credit card resource type tree from wallet application.}
    \label{fig:case-study-resource-trees}
\end{figure}

We demonstrate \toolname{}'s access control enforcement through two case studies across both API-based and browser-based agents. The first case study uses a tic-tac-toe game to illustrate the end-to-end operation of \toolname{}'s core components. The second case study demonstrates enforcement on a realistic flight-booking task involving real-world applications including Outlook Calendar, Expedia, and a payment wallet. Figure~\ref{fig:case-study-resource-trees} shows the resource type trees for the applications used in these case studies.

\subsection{Design Demonstration: Tic-tac-toe Game Example}
\label{sec:tictactoe-example}

In this section, we demonstrate the end-to-end operation of \toolname{} using a simple tic-tac-toe game. We assume the agent can play tic-tac-toe however it likes. The resource we protect is the \emph{history of prior games}, which the agent should not be able to view or delete without authorization. We first show how \toolname{} enforces access control for an API-based agent, then show how the same protection applies to a browser-based agent interacting with the game's web page.

\subsubsection{Application-Defined Components}

To enable any web page or application to work with \toolname{}, the application must provide four components: one or more resource type trees, a set of actions, a resource\_difference function, and a permission function. We walk through each for the tic-tac-toe game.

\paragraph{Resource type tree} We want to implement access control over game history data to prevent agents from viewing or deleting it without proper authorization. We model the resource type tree as a single root node \textit{GameId}, as shown in Figure~\ref{fig:case-study-resource-trees}(a). While we could create more granular child nodes such as \textit{GameStatus} or \textit{GameResult} for finer access control (e.g., preventing access to games the player lost), we keep the model simple with just \textit{GameId}.

\paragraph{Actions} We define two actions for this application: \texttt{read} for viewing game data and \texttt{write} for deleting games.

\paragraph{Resource difference function} The application provides a resource\_difference function for the \textit{GameId} resource type node. Given $\mathrm{resource\_difference}(\textit{GameId}(x), \textit{GameId}(y))$, the function returns an empty set if $y = \texttt{?}$ or $y = x$, and returns a one-element set $\{\textit{GameId}(x)\}$ otherwise. This implementation is conservative: if a single game exists and the user has granted access to it, a precise implementation would recognize that \textit{GameId}(\texttt{?}) is satisfied, but our function returns the full \textit{need} because it does not track the set of existing games. This conservatism is sound (Section~\ref{sec:residual}) and simplifies the implementation.

\paragraph{Permission function (API)} The permission function maps API endpoints to resource value specifications and actions. Consider get\_games() and delete\_game(\textit{game\_id}). The permission function maps get\_games() to \textit{GameId}(\texttt{?}) with read action because it returns all games. For delete\_game(\textit{game\_id}), it maps to \textit{GameId}(\textit{game\_id}), where \textit{game\_id} is resolved at runtime. For instance, if delete\_game is called with ID \textit{45}, the mapped resource value specification becomes \textit{GameId}(``45''), with write action since deletion modifies game data. The API permission function is shown in Figure~\ref{fig:tictactoe-api-mapping}.

\begin{figure}[h]
    \begin{minipage}{\linewidth}
    \begin{minted}[fontsize=\small, frame=single]{python}
def generate_attributes(self, kwargs, endpoint_name, wildcard):
    game_id = kwargs.get('game_id', '?')
    if wildcard:
        game_id = '?'
    
    if endpoint_name == 'get_games':
        resource = f"Game:GameId(?)"
    elif endpoint_name in ['get_game', 'delete_game']:
        resource = f"Game:GameId({game_id})"
    else:
        resource = f"Game:GameId({game_id})"

    action = 'Write' if 'delete' in endpoint_name else 'Read'

    return {
        'resource': resource,
        'action': action
    }
    \end{minted}
    \end{minipage}
    \caption{API permission function for the tic-tac-toe game, mapping endpoints and their arguments to resource value specifications and actions.}
    \label{fig:tictactoe-api-mapping}
\end{figure}

\paragraph{Permission function (web page)} For the game's web interface, the permission function uses a configuration file that maps HTML elements to resource value specifications and actions. For simplicity, we map the game history element to \textit{GameId}(\texttt{?}) with read action, and delete buttons to \textit{GameId}(\texttt{?}) with write action instead of creating separate mappings for individual game IDs. The web element mapping configuration is shown in Figure~\ref{fig:tictactoe-web-mapping}.

\begin{figure}[h]
    \begin{minipage}{\linewidth}
    \begin{minted}[fontsize=\small, frame=single]{json}
{
    "read": [
        "#historyList"
    ],
    "write": [
        "button:contains('Delete')"
    ],
    "Game:GameId(?)": [
        "#historyList",
        "button:contains('Delete')"
    ]
}
    \end{minted}
    \end{minipage}
    \caption{Web element mapping configuration for the tic-tac-toe game, associating CSS selectors with resource value specifications and actions.}
    \label{fig:tictactoe-web-mapping}
\end{figure}

\subsubsection{API-Based Agent Enforcement}

\begin{figure}[h]
    \centering
    \includegraphics[width=0.95\linewidth]{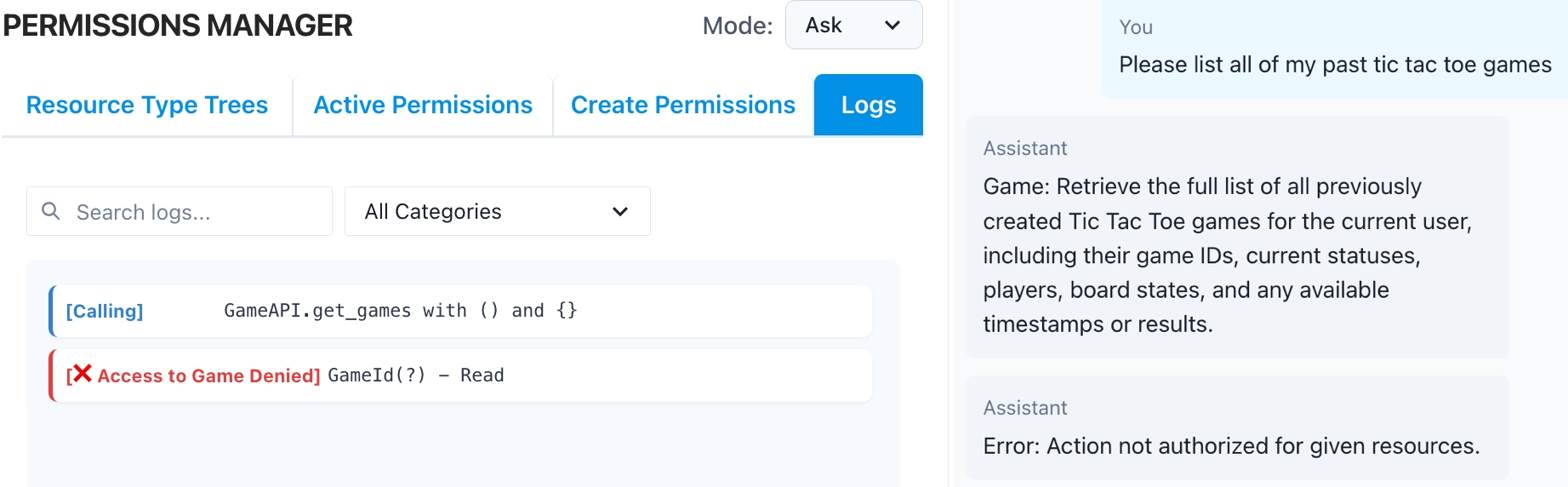}
    \caption{No permissions active. (Left) API access log showing the sufficient permissions and access denial for get\_games. (Right) Chat excerpt showing the permission error returned to the agent.}
    \label{fig:tictactoe-api-log-denied}
\end{figure}

When the agent calls get\_games(), \toolname{} intercepts the call and the permission function returns the sufficient permissions: \textit{GameId}(\texttt{?}) with read action. The permission checking algorithm uses the resource\_difference function to compute \textit{Remaining}. With no active permissions, \textit{Remaining} is not empty and access is denied (Figure~\ref{fig:tictactoe-api-log-denied}).

\begin{figure}[h]
    \centering
    \includegraphics[clip, trim=0cm 7.7cm 0cm 0cm,width=0.98\linewidth]{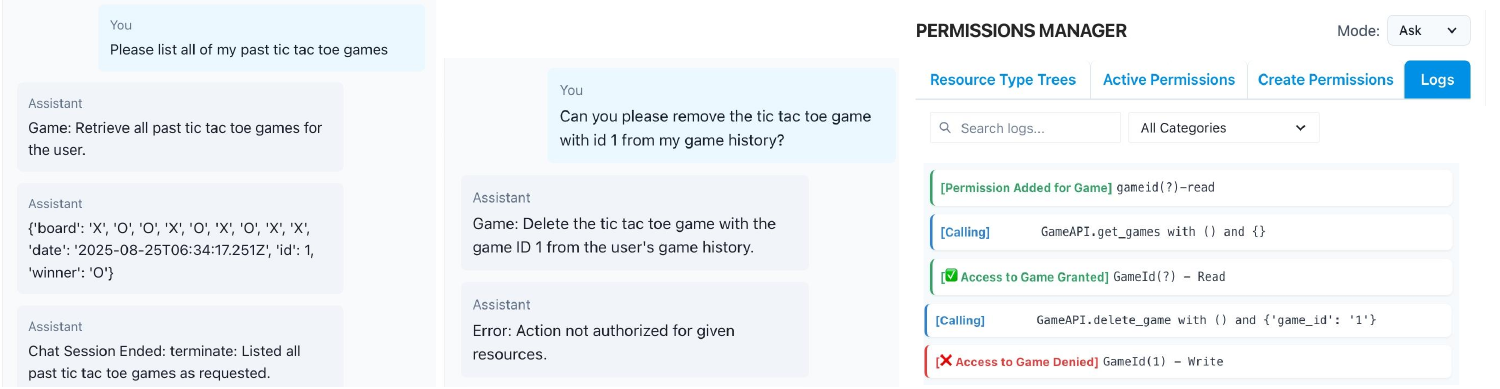}
    \caption{Read-only permissions active. (Left) Chat excerpt: agent successfully reads game data. (Center) Chat excerpt: agent fails to delete game history due to missing write permission. (Right) API access log: get\_games succeeds, delete\_game is denied.}
    \label{fig:tictactoe-api-log-conflict}
\end{figure}

\begin{figure}[h]
    \centering
    \includegraphics[clip, trim=0cm 9cm 7cm 0cm,width=\linewidth]{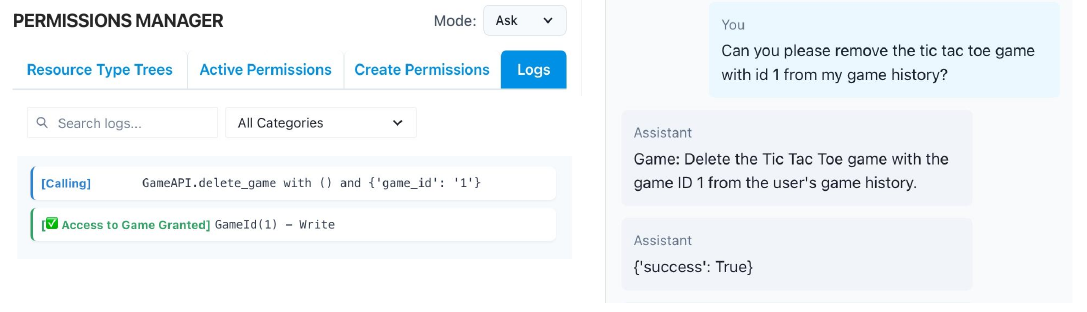}
    \caption{Read and write permissions active. (Left) Chat excerpt: agent successfully deletes game data. (Right) API access log: delete\_game is granted.}
    \label{fig:tictactoe-api-log-granted}
\end{figure}

After granting read permission, get\_games() succeeds. However, when the agent attempts to delete a game, \toolname{} intercepts delete\_game() and the permission function returns write access as the sufficient permission. With no active write permission, the request is denied (Figure~\ref{fig:tictactoe-api-log-conflict}). When both read and write permissions are active, the agent can successfully delete games (Figure~\ref{fig:tictactoe-api-log-granted}).

\subsubsection{Browser-Based Agent Enforcement}

For the game's web interface, enforcement works at the HTML element level. When a browser-based agent is instructed to look up game history, it opens the game web app. \toolname{} intercepts the page before the agent can access it and uses the configuration file to determine the sufficient permissions for each mapped element. The permission checking algorithm computes \textit{Remaining} against the active permissions using the resource\_difference function. Elements for which \textit{Remaining} is not empty are blocked with a solid dark box displaying a \emoji{prohibited} sign, informing the agent of restricted access.

\begin{figure}[h]
    \centering
    \includegraphics[clip, trim=0cm 8.0cm 5.0cm 0cm,width=\linewidth]{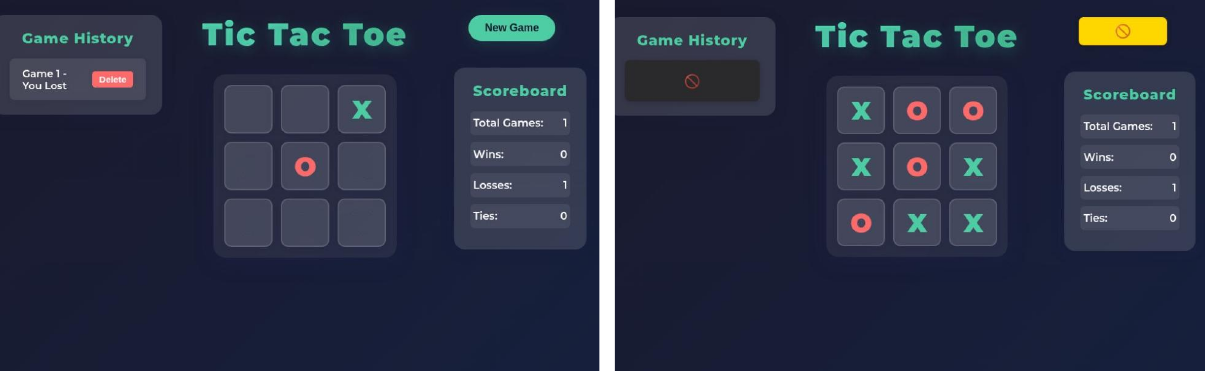}
    \caption{No permissions active. (Left) Original browser view with all elements visible. (Right) View shown to the agent, with game history elements blocked by solid overlays.}
    \label{fig:tictactoe-browser-blocked}
\end{figure}

\begin{figure}[h]
    \centering
    \includegraphics[clip, trim=0cm 8.0cm 0.5cm 0cm,width=\linewidth]{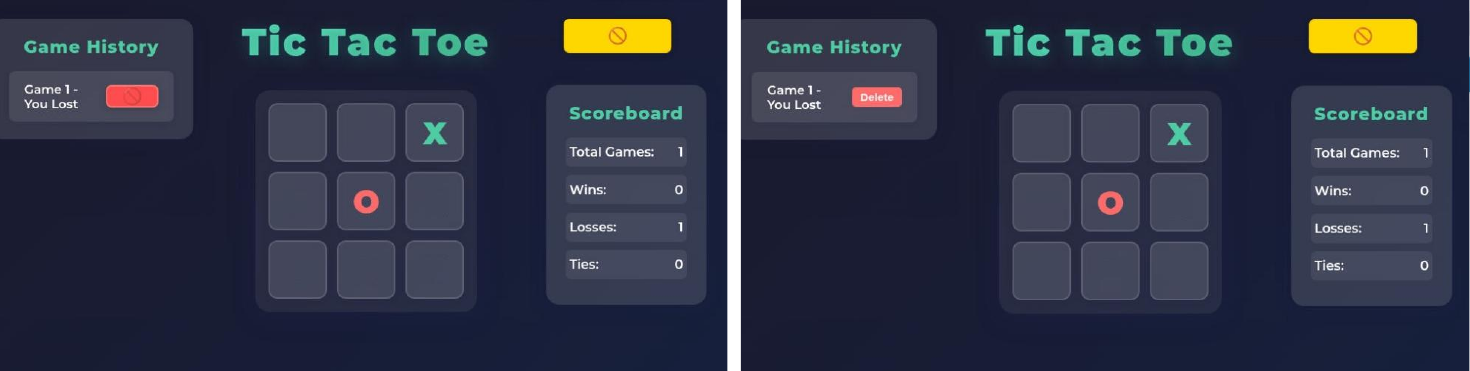}
    \caption{(Left) Read-only permission: game history is visible but delete buttons are blocked. (Right) Read and write permissions: delete buttons become accessible.}
    \label{fig:tictactoe-browser-partial-access}
\end{figure}

With no active permissions, all game history elements are blocked (Figure~\ref{fig:tictactoe-browser-blocked}, right). After granting read permission for \textit{GameId}(\texttt{?}), the game history becomes visible, but delete buttons remain blocked because no write permission exists (Figure~\ref{fig:tictactoe-browser-partial-access}, left). With both read and write permissions active, the delete buttons become accessible (Figure~\ref{fig:tictactoe-browser-partial-access}, right).

This example demonstrates \toolname{} with both an API and a web page. Real-world applications are more complex, with multiple elements connected to various resource types. We have scaled this technique to realistic use cases such as Outlook Calendar and Expedia.

\subsection{Flight Booking Demonstration}

We demonstrate \toolname{} on a realistic, multi-step flight-booking task that involves three applications: a calendar (Outlook), a travel service (Expedia), and a payment wallet. We decompose the task into sub-tasks that test access control boundaries. From the perspective of \toolname{}, accidental and malicious overreaches are indistinguishable and handled identically by the access control layer. The demonstration begins with no active permissions granted, and we show enforcement for both API-based and browser-based agents using the same underlying permissions.

Each application provides the four components required by \toolname{} (Section~\ref{sec:tictactoe-example}). The calendar uses the resource type tree from Figure~\ref{fig:resource-examples}(a), with Year, Month, and Day resource type nodes, and defines read, write, and create actions. Its permission function maps reading and availability-checking endpoints to read access and reservation endpoints to create access, deriving resource value specifications from the \texttt{start\_time} and \texttt{duration} arguments. Expedia's resource type tree is shown in Figure~\ref{fig:case-study-resource-trees}(b). Its permission function maps search and info endpoints to read access on flight data, and booking endpoints to create access on the specific flight being booked (e.g., Flight(``DL 1847'')). The wallet defines multiple resource type trees, one of which is the credit card tree from Figure~\ref{fig:resource-examples}(c). The wallet's permission function maps query endpoints to read access.

\subsubsection{API-Based Agent Enforcement}

\begin{figure}[h]
  \centering
  \includegraphics[width=\linewidth]{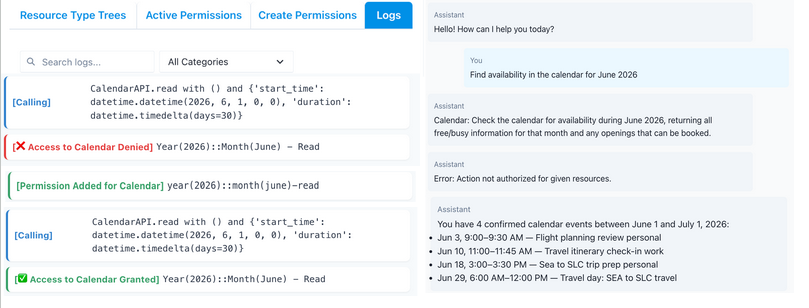}
  \caption{API-based agent enforcement for Scenario A: calendar read for June 2026 blocked, then granted.}
  \label{fig:api-transcripts-aa}
\end{figure}

\paragraph{Scenario A (Unauthorized Calendar Read)}
The user asks the agent to find calendar availability for June 2026. \toolname{} intercepts the resulting API call, and the permission function returns the sufficient permissions (Year(2026)\texttt{::}Month(June), read). Because no active permission covers this need, the permission checking algorithm produces a nonempty \textit{Remaining} set, and the call is blocked. Once the user grants the appropriate permission through the dashboard, the agent retries the same call successfully. Chat excerpts and the \toolname{} access log for this scenario appear in Figure~\ref{fig:api-transcripts-aa}.

\begin{figure}[h]
  \centering
  \includegraphics[width=\linewidth]{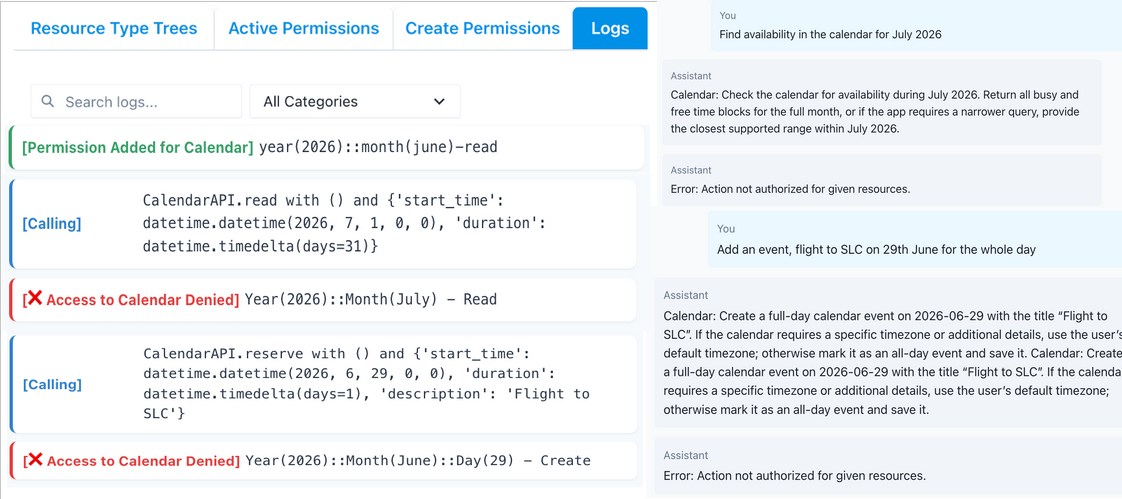}
  \caption{Scenario B: July 2026 read blocked (permission covers only June); create for 29 June 2026 blocked (read-only permission).}
  \label{fig:api-transcripts-bb}
\end{figure}

\paragraph{Scenario B (Scope and Action Boundaries)}
With the June 2026 read permission now active, this scenario exercises two orthogonal boundaries: resource scope and action type. First, the user requests availability for July 2026. The permission function returns Year(2026)\texttt{::}Month(July) as the required resource, but the permission checking algorithm determines that the active permission for Month(June) does not cover Month(July), so the call is blocked. Second, a create action (adding an event on 29 June 2026) is blocked because the active permission grants only read access, not create. Both cases are shown in Figure~\ref{fig:api-transcripts-bb}.

\begin{figure}[h]
  \centering
  \includegraphics[width=\linewidth]{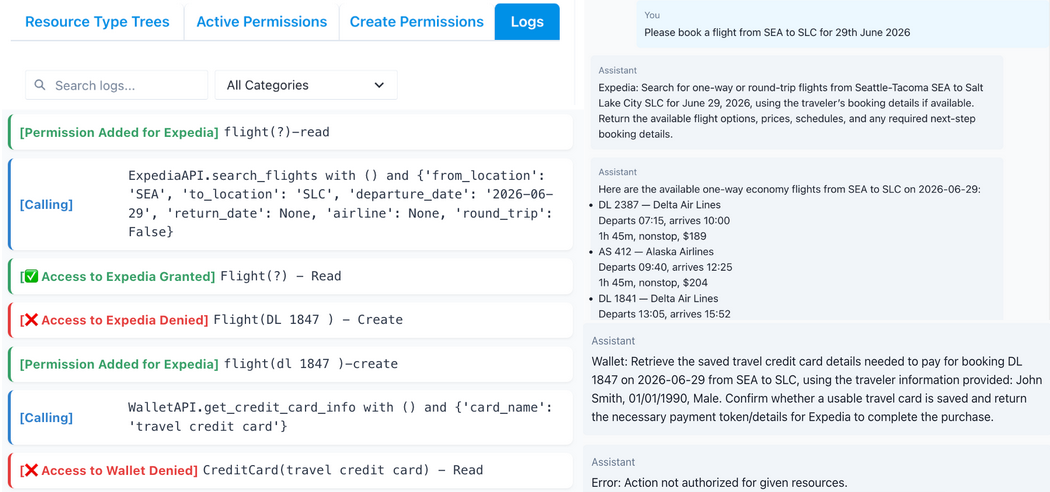}
  \caption{Scenario C: flight read allowed; create booking for DL 1847 blocked until create permission added; credit card read denied.}
  \label{fig:api-transcripts-cc}
\end{figure}

\paragraph{Scenario C (Flight Booking and Cross-Application Access)}
The user instructs the agent to book a flight for 29 June 2026 based on the retrieved calendar availability. The user has already granted read access to all flight data, so the agent can search and view flight options. When the user asks the agent to book a specific flight (e.g., DL 1847), the agent attempts to create the booking. The permission function returns the sufficient permission (Flight(``DL 1847''), create), but no active permission covers it, so the request is denied. Once the user adds create permission for that flight through the dashboard, the agent retries and the booking proceeds. The agent then attempts to retrieve credit card information from the wallet for payment. \toolname{} intercepts the call; the permission function returns the sufficient permissions for the credit card resource, but the permission checking algorithm finds no active read permission covering it, so the request is denied. Figure~\ref{fig:api-transcripts-cc} shows the chat excerpts and access log for this scenario.

In the scenarios above, we manually triggered requests that should be blocked (e.g., requesting July when only June was permitted) to demonstrate how \toolname{} handles unauthorized access attempts, whether caused by agent bugs or malicious intent. We also created permissions manually through the dashboard. In practice, permissions can also be generated from natural language descriptions or inferred automatically using the handling modes described in Section~\ref{sec:handling-modes}.

\subsubsection{Browser-Based Agent Enforcement}

We replicate the same workflow using a browser-based agent to demonstrate that \toolname{} enforces the same permissions at the UI level. Rather than blocking API payloads, \toolname{} injects CSS overlays to visually block unauthorized DOM elements before the agent can access them.

\begin{figure}[ht!]
  \centering
  \footnotesize
  \begin{subfigure}[t]{0.32\linewidth}
    \centering
    \includegraphics[width=\linewidth,trim=0 0 0 0,clip]{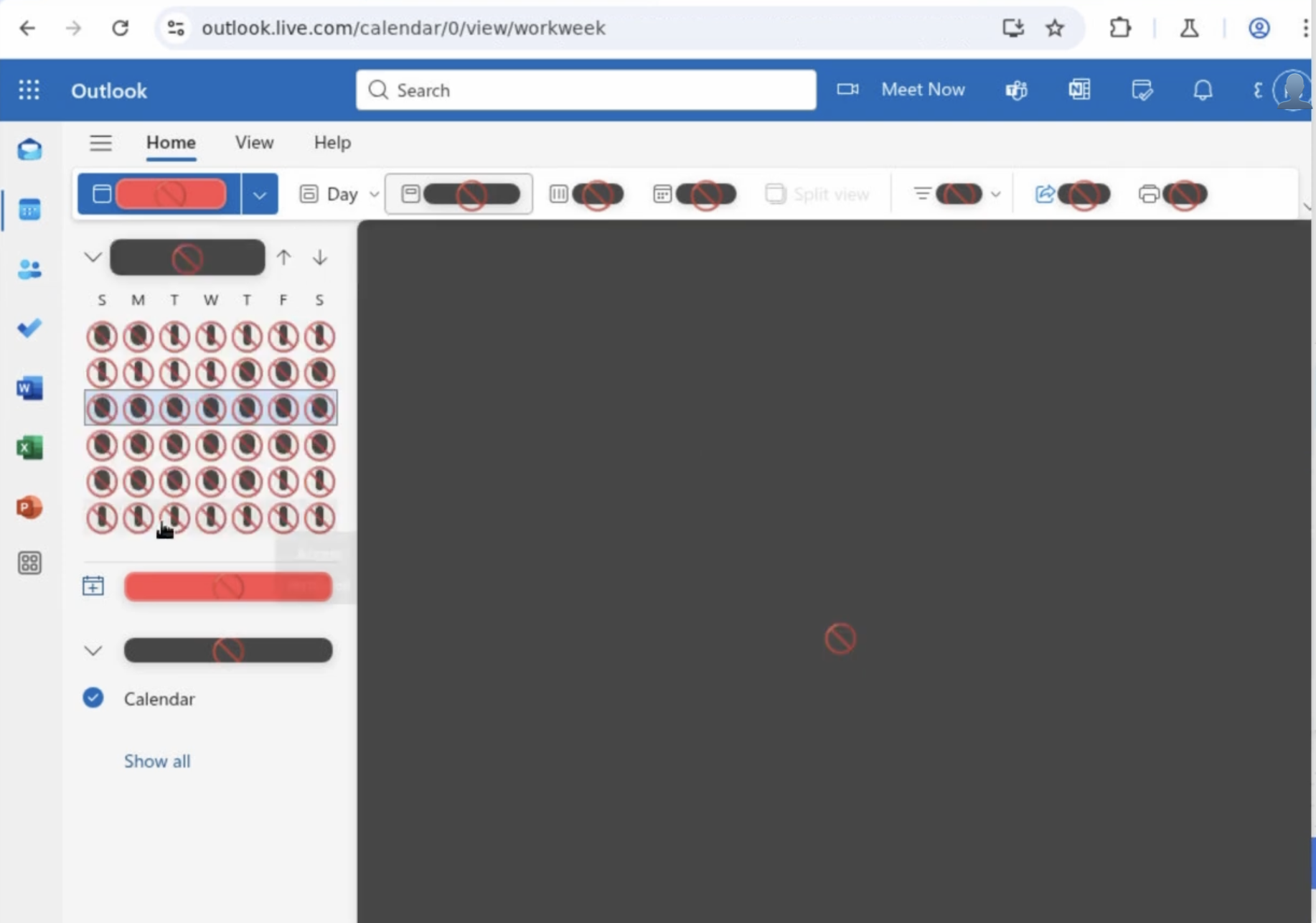}
    \caption{Outlook Calendar with no permissions. The calendar grid and date cells are blocked.}
    \label{fig:browser-transcripts-a}
  \end{subfigure}\hfill
  \begin{subfigure}[t]{0.32\linewidth}
    \centering
    \includegraphics[width=\linewidth,trim=0 0 0 0,clip]{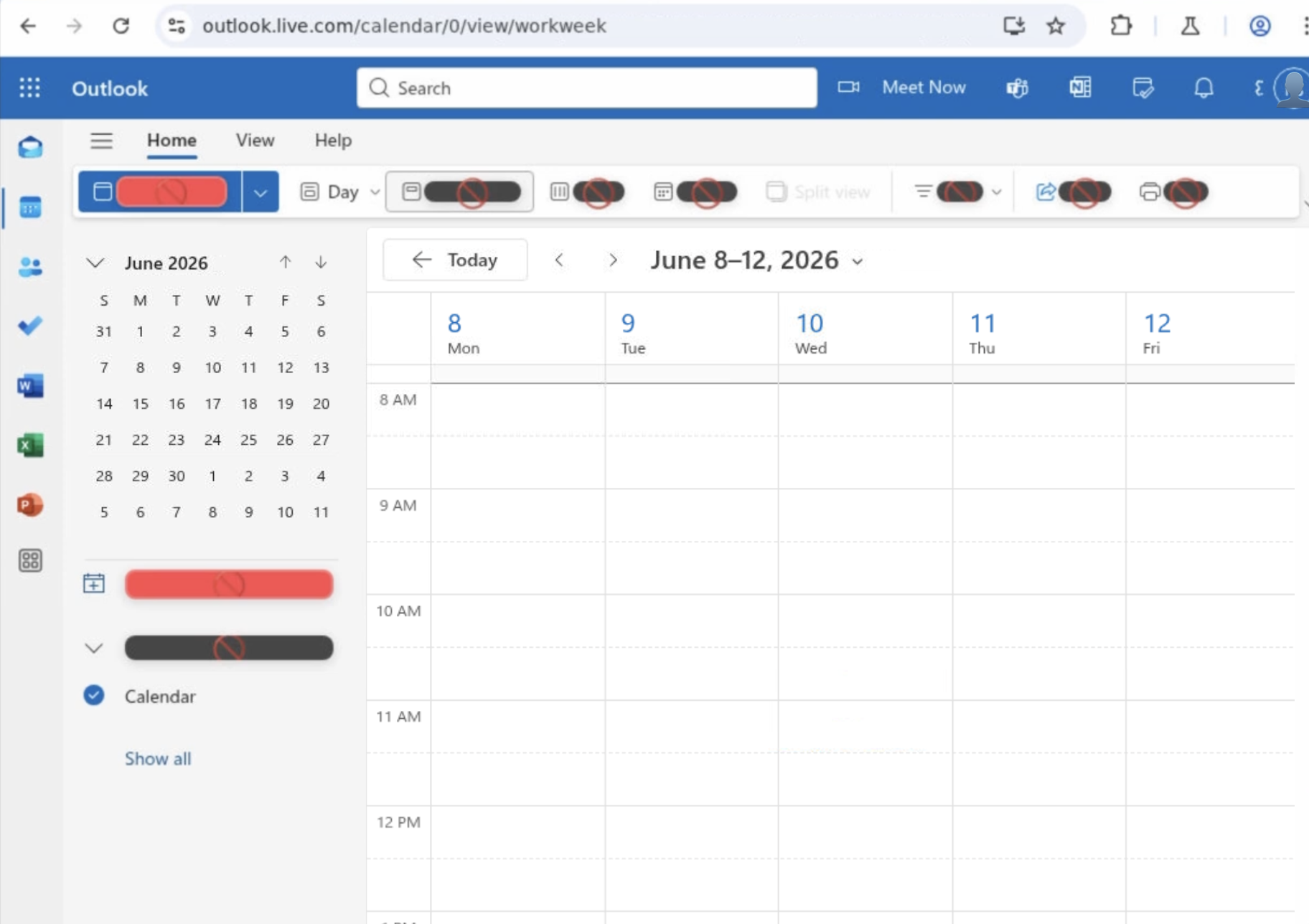}
    \caption{After granting read for June 2026. The week view is now visible.}
    \label{fig:browser-transcripts-b}
  \end{subfigure}\hfill
  \begin{subfigure}[t]{0.32\linewidth}
    \centering
    \includegraphics[width=\linewidth,trim=0 0 0 0,clip]{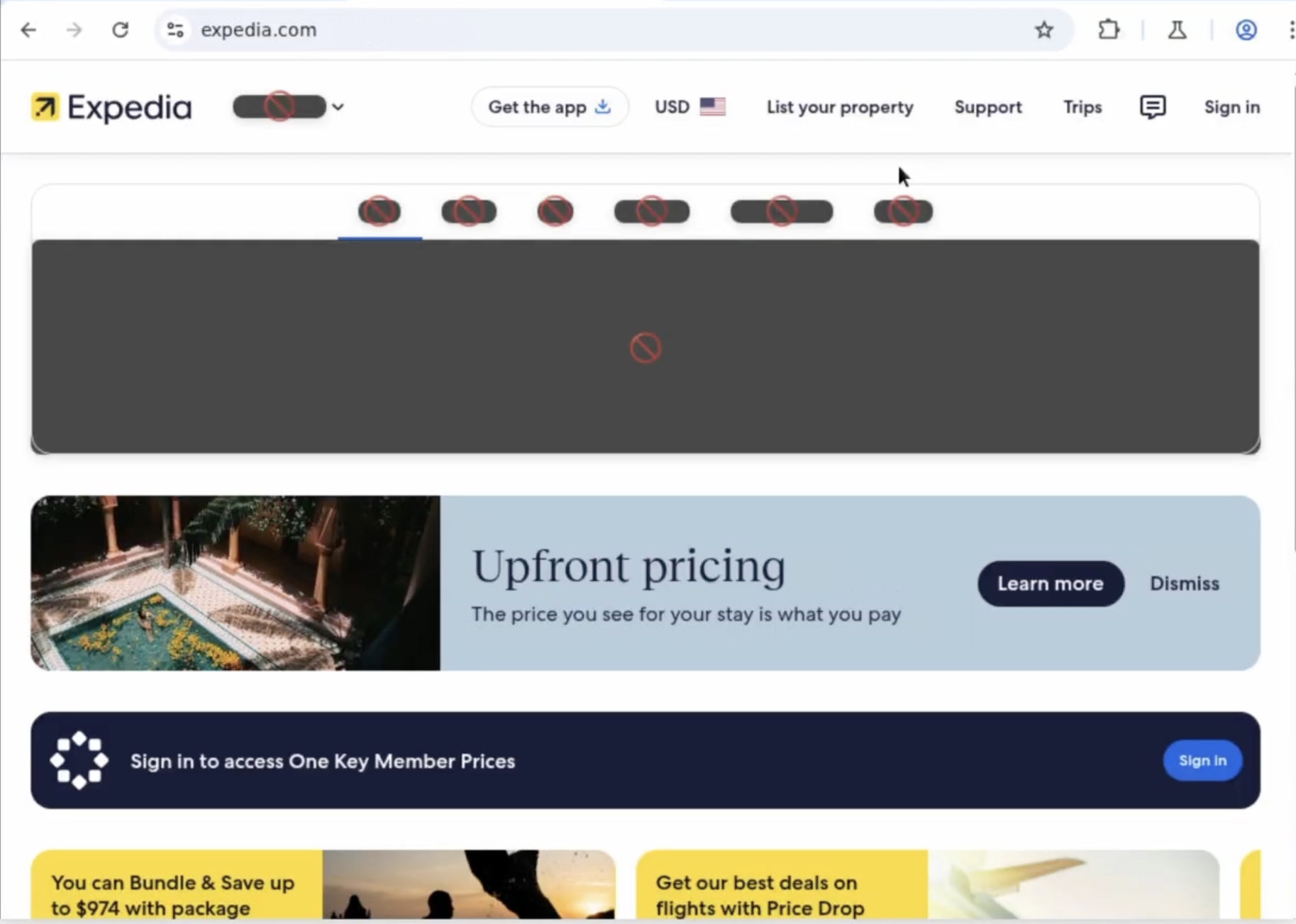}
    \caption{Expedia with no permissions. The search area is blocked.}
    \label{fig:browser-transcripts-c}
  \end{subfigure}\\[0.5em]
  \begin{subfigure}[t]{0.32\linewidth}
    \centering
    \includegraphics[width=\linewidth,trim=0 0 0 0,clip]{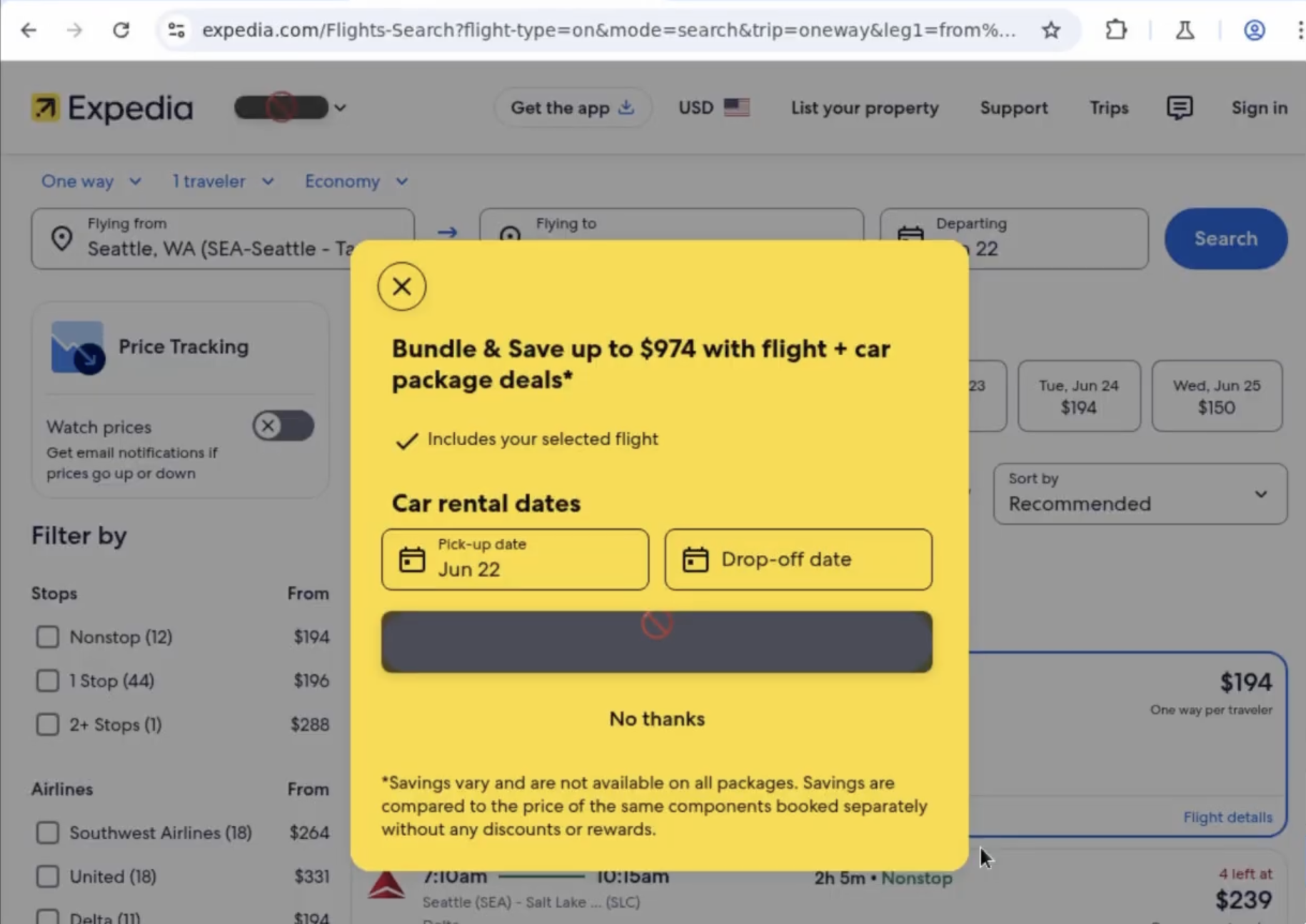}
    \caption{Car rental popup remains blocked despite flight read permission.}
    \label{fig:browser-transcripts-d}
  \end{subfigure}\hfill
  \begin{subfigure}[t]{0.32\linewidth}
    \centering
    \includegraphics[width=\linewidth,trim=0 0 0 0,clip]{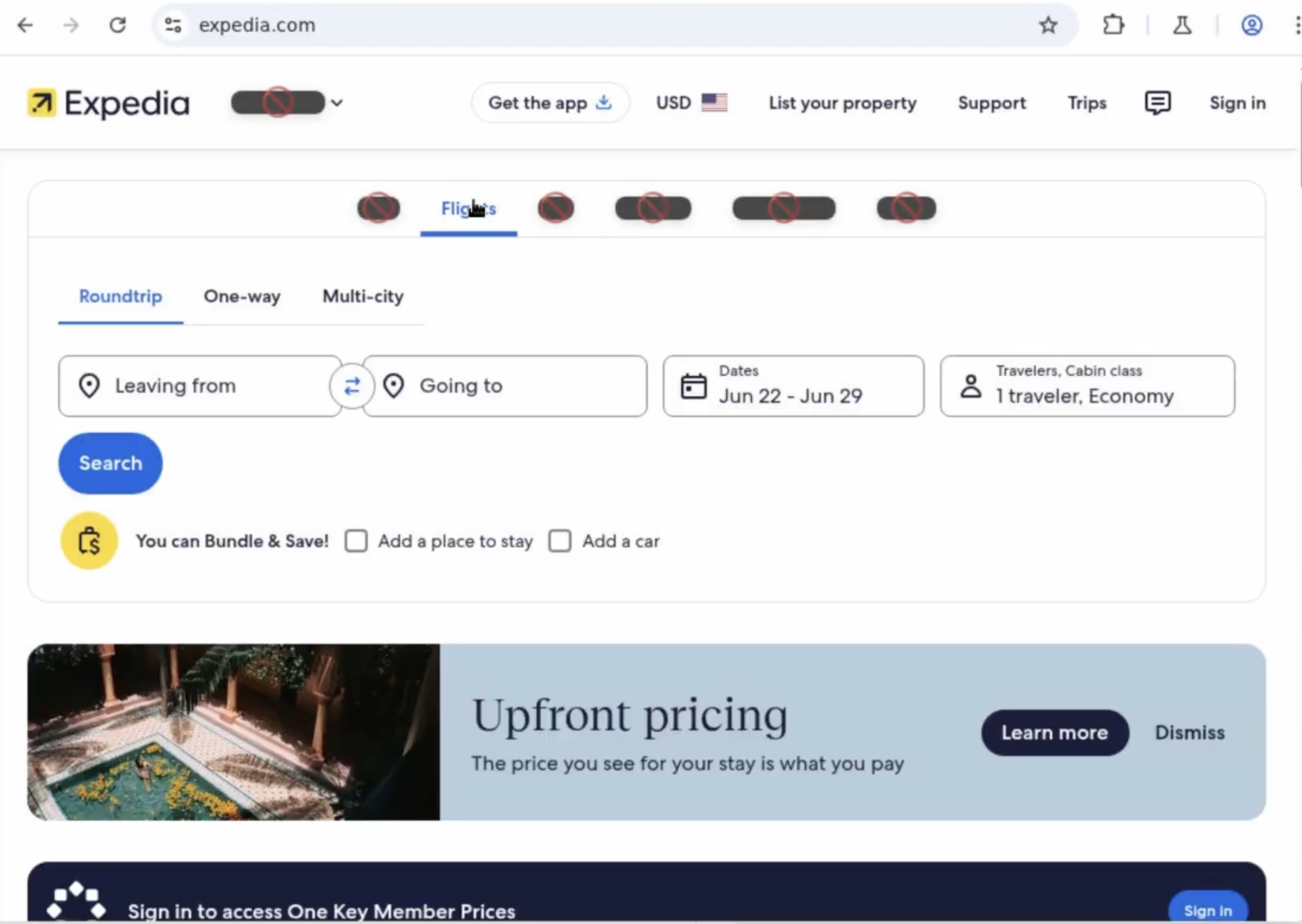}
    \caption{After granting flight read. Search form and flight results are visible.}
    \label{fig:browser-transcripts-e}
  \end{subfigure}\hfill
  \begin{subfigure}[t]{0.32\linewidth}
    \centering
    \includegraphics[width=\linewidth,trim=0 0 0 0,clip]{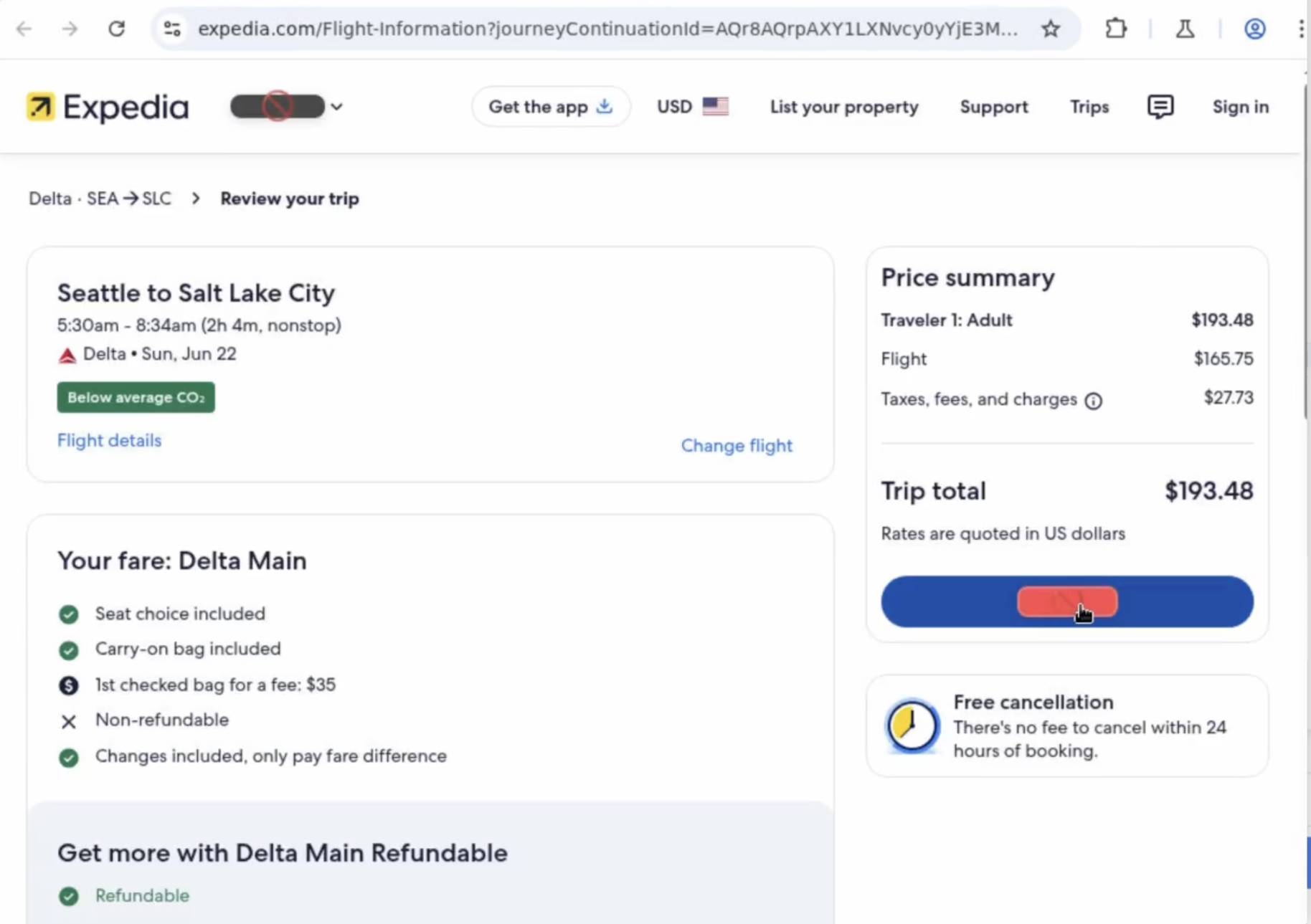}
    \caption{After create for flight granted, booking proceeds; credit card read blocked until user grants read permission.}
    \label{fig:browser-transcripts-f}
  \end{subfigure}
  \caption{Browser-based agent enforcement. \toolname{} injects CSS overlays to block unauthorized elements before the agent can access them.}
  \label{fig:browser-transcripts}
\end{figure}

\paragraph{Scenario A (Calendar Blocking and Incremental Grant)}
The agent is instructed to open the calendar at \textit{outlook.live.com/calendar}. Before the agent can access the page, \toolname{} intercepts it and the configuration file determines the sufficient permissions for each mapped UI element. With no active permissions, the permission checking algorithm finds all calendar elements unauthorized, and CSS is injected to block all mapped resources (Figure~\ref{fig:browser-transcripts-a}). The agent reports that it cannot view the schedule. After the user grants read access for June 2026, the calendar grid becomes visible and the agent parses the availability (Figure~\ref{fig:browser-transcripts-b}).

We then test two additional boundaries. First, requesting July 2026 leaves the calendar blocked because the active permission covers only June. Second, requesting to add an event on 29 June 2026 reveals that the event-creation UI, including the save button, remains blocked because the active permission grants only read access, not create.

\paragraph{Scenario B (Expedia Flight Booking)}
The agent navigates to Expedia to book a flight (SEA to SLC, 29 June 2026). With no Expedia permission, the flight search area is visually blocked (Figure~\ref{fig:browser-transcripts-c}). After the user grants read access to flight resources, flight content and search controls become visible (Figure~\ref{fig:browser-transcripts-e}), while elements outside the granted scope, such as car rental pop-ups, remain blocked because the permission covers only flight resources (Figure~\ref{fig:browser-transcripts-d}). When the agent selects a flight (e.g., DL 1847) and attempts to create the booking, that action is blocked until the user grants create permission for that specific flight. Once the booking proceeds, the agent attempts to retrieve credit card information for payment; that access is blocked until the user grants read permission for the credit card resource (Figure~\ref{fig:browser-transcripts-f}).

Together, the API and browser scenarios demonstrate that \toolname{} successfully enforces access control regardless of whether the agent interacts through API calls or front-end DOM manipulation, using the same permission representation and the same set of active permissions for both.

\section{Limitations and Future Work}
\label{sec:limit}

\noindent\textbf{Mapping File Requirements and Maintenance.}
The permission function for APIs and the configuration file for web pages must be written and maintained. For APIs, the natural provider is the application developer, who understands the semantics of each endpoint. For web pages, configuration files can also be developed by third parties or even a community effort, since one does not need server access to provide a configuration for an online web application. For example, we were able to create a working configuration for Outlook Calendar without any involvement from Microsoft. However, third-party configurations are fragile. If the application changes how the DOM is represented, the configuration can become out of date. Writing these mappings can be challenging, and they require ongoing maintenance to stay consistent with evolving web pages and API endpoints. Errors in mappings can lead to over-restriction (blocking legitimate access) or under-restriction (allowing unauthorized access) in permission enforcement. We consider this the most significant practical limitation of \toolname{}. Developing tools to validate permission functions and configuration files, and providing guidance for writing correct mappings, is an important direction for future work.

\noindent\textbf{Complexity in Browser Mapping Definitions.}
The current mapping language for browsers is expressive enough to handle the applications we have tested, but it could be improved by supporting additional programming constructs such as conditional statements dependent on page content and iteration over dynamically generated child elements. We plan to make the mapping language more expressive in future work.

\noindent\textbf{Modeling Complex Access Control Requirements.}
Certain access control requirements are difficult to express in the current framework. For instance, allowing checkout access only when the amount is less than 200 USD would require modeling \textit{checkout} and \textit{checkout under 200 USD} as separate resource type nodes. In future work, we plan to add conditionals to permissions that allow comparisons on resource values, enabling permissions such as \textit{access to checkout with amount less than 200 USD}.

\noindent\textbf{Finding Good Permissions.}
\toolname{} focuses exclusively on enforcing permissions, not on determining what the right permissions should be. While it provides methods for generating permissions (manually, via LLM, or inferred from user input), the question of what constitutes good permissions for a given task remains open. This separation makes \toolname{} a useful testbed for evaluating methods that generate permissions. For example, user-driven access control~\cite{udac} infers implicit permissions from user intent. \toolname{} can evaluate the effectiveness of such methods by checking whether the generated permissions correctly allow intended accesses and block unintended ones.

\noindent\textbf{Permission Lifecycle Management.}
Permissions in \toolname{} persist until explicitly removed. This creates a risk. A user may grant write permission on a calendar date to modify an event, and the agent could later use the same permission to delete a different event on that date. Single-use permissions that expire after one access would mitigate this, but raise questions about who decides when permissions should expire and how to handle cases where multiple permissions cover overlapping resources. Permission lifecycle management falls under access control policy, which \toolname{} does not currently address.

\section{Related Work}
\label{sec:related}

Agents bring new challenges to access control. Existing techniques have influenced the design of \toolname{} but are not directly applicable. We first discuss traditional access control systems, then agent containment and information flow control, and finally access control specifically designed for agents.

\noindent\textbf{Traditional Access Control Systems.}
Traditional access control has evolved through several models~\cite{access-control-survey}. Unix systems established file-based access control through permission bits (read/write/execute for user/group/others)~\cite{unix-access-control, unix-privilege}. \toolname{}'s resource type trees draw directly from the Unix file hierarchy, pairing hierarchical resource organization with application-defined actions. Unlike Unix, where a user can have permission on a parent directory but not on a child file, \toolname{} currently ensures that permission on a parent resource type node implies permission on all children.

Mobile platforms and web browsers extended the Unix model through declarative permissions that specify required capabilities upfront~\cite{mobile-permission-survey}. Android introduced manifest-based declarations with protection levels and runtime consent for dangerous permissions~\cite{android-permission-system-design}. Web browsers implement origin-based permission scoping with explicit user consent for capabilities such as geolocation, camera, and microphone~\cite{web-permission-api, contego}. Like these systems, \toolname{} computes what permissions are required. Unlike them, it does so at runtime based on the agent's actual interactions rather than through static declarations.

Role-Based Access Control (RBAC) organizes permissions through role definitions with hierarchical inheritance and separation of duties~\cite{RBAC}. Attribute-Based Access Control (ABAC) evaluates permissions dynamically based on subject, object, and environment attributes~\cite{ABAC}. \toolname{}'s permission model uses hierarchical resources (a permission on a parent resource type node covers all children) and dynamic evaluation (sufficient permissions are computed at runtime from actual argument values, as in ABAC).

Zanzibar~\cite{zanzibar} introduced a unified authorization system across Google's services by representing all permissions as relation tuples in a global graph (Relation-Based Access Control). Just as Zanzibar decoupled access control from individual microservices to provide a universal vocabulary, \toolname{} decouples access control from individual agent implementations. However, while Zanzibar evaluates permissions by traversing relationship graphs between users and objects, \toolname{} evaluates agent interactions against hierarchical resource type trees, a structure suited to DOM and API-based agents.

\noindent\textbf{Agent Containment and Information Flow Control.}
Several frameworks mitigate indirect prompt injection and unauthorized operations by containing untrusted external inputs. IsolateGPT~\cite{isolate-gpt} runs individual tools in separate system-level sandboxes, mediating cross-app interactions through a central planner. F-Secure~\cite{f-secure} disaggregates the agent into a context-aware pipeline where a security monitor filters untrusted inputs before they reach the planning process. These systems provide execution isolation but do not manage fine-grained permissions over what resources the agent is allowed to access.

A parallel line of work secures agents by tracking and separating control and data flows. CAMEL~\cite{camel} uses a dual-LLM architecture with taint tracking to ensure untrusted data cannot hijack control flow. ACE~\cite{ace} decouples planning into abstract and concrete phases, using static analysis to verify that execution plans satisfy information-flow constraints. FIDES~\cite{fides} attaches confidentiality and integrity labels to all processed data, introducing primitives to selectively hide or inspect tainted variables.

These approaches treat agent security as a data containment problem by tracking how untrusted data moves through an agent to prevent it from influencing trusted operations. \toolname{} cannot mitigate these threats directly, but access control serves as an essential component that enables other security mechanisms.

\noindent\textbf{Access Control for AI Agents.}
Progent~\cite{progent} and Conseca~\cite{conseca} are closest to \toolname{} in providing access control specifically for agents. They differ in how policies are defined. Progent enforces the principle of least privilege at the tool-call level through a JSON-based domain-specific language that lets developers specify which tools are allowed, forbidden, or conditionally permitted based on their arguments (e.g., restricting an email tool to specific recipients). A deterministic runtime evaluates each tool call against this policy. Conseca takes a context-aware approach. A separate Policy Generator LLM, given only trusted context (user intent, system state), produces a task-specific security policy for each request. Untrusted context (e.g., web content) is isolated from policy generation, and a deterministic Policy Enforcer evaluates the agent's tool calls against this temporary policy before execution.

\toolname{} differs from both by modeling permissions over \emph{resources} rather than over tool calls or task contexts. While Progent can constrain tool arguments (e.g., restricting email recipients), these constraints are specific to each tool. \toolname{} defines the permission once on the resource itself, and enforcement applies uniformly regardless of how the resource is accessed.

\section{Conclusion}
\label{sec:conclusion}

Traditional software systems treat access control as foundational infrastructure. It is unimaginable to deploy a multi-user operating system without file permissions or a web application without authentication. Yet today's LLM agents, which autonomously call APIs and navigate web pages on behalf of users, operate with no comparable mechanism. In this paper, we introduced \toolname{}, a framework that brings fine-grained, resource-centric access control to agents. The key insight behind \toolname{} is that permissions should be defined over \textit{resources} rather than over tool calls, and that representing resources as hierarchical types yields a permission model that is uniformly enforceable across API-based and browser-based agents. Our case studies on real-world applications such as Outlook Calendar and Expedia demonstrate that agents can operate within restricted permissions while being prevented from overstepping their authorization.

\toolname{} is intentionally scoped to enforcement. It ensures that a given set of permissions is respected but does not prescribe what those permissions should be. This separation is both a practical design choice and a research opportunity. From a practical standpoint, \toolname{} can be deployed today with manually authored permission functions and configuration files, as our case studies demonstrate. At the same time, this separation opens a research agenda around the ecosystem built on top of an enforcement layer: What constitutes a \textit{good} set of permissions for a task? How should permissions evolve over the course of a session? Can the mapping between application interfaces and resource types be learned or crowdsourced rather than manually authored? We believe that establishing a principled enforcement foundation, as \toolname{} does, is a prerequisite for meaningfully investigating these questions, just as access control frameworks enabled decades of research on access control policies, auditing, and privilege separation.

\appendix

\newglossarystyle{definitionstyle}{%
  \setglossarystyle{list}%
  \renewenvironment{theglossary}{%
    \begin{description}[leftmargin=0pt,labelwidth=0pt,itemindent=0pt,labelsep=0.5em]%
  }{%
    \end{description}%
  }%
  \renewcommand*{\glossentry}[2]{%
    \item[\normalfont\textbf{\glstarget{##1}{##2}}]%
    \glsentrydesc{##1}\glspostdescription%
    \vspace{0.5\baselineskip}%
  }%
  \renewcommand*{\subglossentry}[3]{%
    \glossentry{##2}{##3}%
  }%
  \renewcommand*{\glsgroupskip}{%
    \vspace{0.3\baselineskip}%
  }%
  \renewcommand*{\glossaryheader}{}%
  \renewcommand*{\glsgroupheading}[1]{}%
}

\bibliographystyle{plain} %
\bibliography{main} %

\end{document}